\documentclass[twocolumn,pra]{revtex4-1}
\usepackage{graphicx}\usepackage{braket}
\usepackage{amsmath,amsfonts,amssymb,amstext,amscd,amsthm}
\usepackage[colorlinks, citecolor=blue]{hyperref}
\begin{document}	
\title{Quantum walker as a probe for its coin parameter}
\author{Shivani Singh}
\email{shivanis@imsc.res.in}
\affiliation{The Institute of Mathematical Sciences, 
C. I. T, campus, Taramani, Chennai, 600113, India}
\affiliation{Homi Bhabha National Institute, Training 
School Complex, Anushakti Nagar, Mumbai 400094, India}
\author{C. M. Chandrashekar}
\email{chandru@imsc.res.in}
\affiliation{The Institute of Mathematical Sciences, 
C. I. T, campus, Taramani, Chennai, 600113, India}
\affiliation{Homi Bhabha National Institute, Training 
School Complex, Anushakti Nagar, Mumbai 400094, India}
\author{Matteo G. A. Paris}
\email{matteo.paris@fisica.unimi.it}
\affiliation{Quantum Technology Lab, Dipartimento di Fisica 
{\em Aldo Pontremoli}, Universit\`a degli Studi di Milano, 
I-20133 Milano, Italy}
\begin{abstract}
In discrete-time quantum walk (DTQW) the walker's coin space 
entangles with the position space after the very first step 
of the evolution. This phenomenon may be exploited to obtain 
the value of the coin parameter $\theta$ by performing measurements 
on the sole position space of the walker. In this paper, we evaluate
the ultimate quantum limits to precision for this class of 
estimation protocols, and use this result to assess 
measurement schemes having limited access to the position
space of the walker in one dimension. We find that the quantum Fisher information 
(QFI) of the walker's position space $H_w(\theta)$ increases with $\theta$ and 
with time which, in turn, may be seen as a metrological 
resource. We also find a difference in the QFI of {\em bounded} 
and {\em unbounded} DTQWs, and provide an interpretation of the 
different behaviors in terms of interference in the position 
space. Finally, we compare $H_w(\theta)$ to the full 
QFI $H_f(\theta)$, i.e., the QFI of the walkers position 
plus coin state, and find that their ratio is dependent on 
$\theta$, but saturates to a constant value, 
meaning that the walker may probe its coin parameter quite 
faithfully.
\end{abstract}
\maketitle
\section{\label{sec1}Introduction}
Quantum walk is the quantum analog of random walk which, 
in turn, provides a relevant model for the dynamics of 
various classical  systems \cite{GVR1958, feynman1986quantum}. 
Quantum superposition and interference strongly affects 
the dynamics of a quantum walker and this leads to a 
quadratically faster spread in position space when compared to a classical walker \cite{10.2307/3214153, aharonov1993quantum, meyer1996quantum, kempe2003quantum, venegas2012quantum, childs2003exponential, childs2004spatial}. This feature made 
quantum walks a powerful tool in quantum computation \cite{ambainis2007quantum, magniez2007quantum, buhrman2006quantum, farhi2007quantum, konno2008quantum}, 
as well as to model the 
dynamics of different quantum systems, such as energy transport 
in photosynthesis \cite{engel2007evidence, mohseni2008environment}, 
quantum percolation\,\cite{chandrashekar2014quantum,
kollar2012asymptotic}, and graph isomorphism 
\cite{douglas2008classical}. 
\par
As in classical random walk, quantum walk has also 
been developed in two forms, continuous-time and 
discrete-time quantum walk (DTQW). Both the variants have been 
shown to efficiently implement any quantum computational task \cite{childs2009universal,douglas2008classical}.  
Continuous-time quantum walk is defined only on position 
Hilbert space, whereas discrete-time quantum walk is defined 
on a joint position and {\em coin} Hilbert space, thus providing an
additional degree of freedom to control the dynamics. Upon 
tuning the different parameters of the evolution operators 
of DTQW, one may control and engineer the dynamics in order 
to {\em simulate} various quantum phenomena such as localization\,\cite{joye2012dynamical, chandrashekar2012disorder, chandrashekar2015localized}, topological phase\,\cite{obuse2011topological, kitagawa2010exploring}, neutrino oscillation\,\cite{mallick2017neutrino, di2016quantum}, and relativistic quantum dynamics\,\cite{strauch2006relativistic, chandrashekar2010relationship, chandrashekar2013two, di2013quantum, di2014quantum, arrighi2016quantum, perez2016asymptotic}. 
Quantum walks have been experimentally 
implemented in various physical systems such as NMR\,\cite{ryan2005experimental}, photonics\,\cite{schreiber2010photons, broome2010discrete, peruzzo2010quantum, perets2008realization}, 
cold atoms\,\cite{karski2009quantum}, and trapped ions\,\cite{schmitz2009quantum, zahringer2010realization}. 
\par
Evolution in discrete-time quantum walk is defined by 
unitary {\em coin operation} followed by a unitary
{\em position shift operator}. The shift operator evolves 
the walker in a superposition of the position states, 
with amplitudes governed by the operation on coin 
Hilbert space. The most general unitary coin operator in one dimension has 
three independent parameters \cite{chandrashekar2008optimizing} 
and provides an ample control over the dynamics, but already 
one and two parameter coins are extremely useful in simulating 
various physical systems in one dimension. For example, different combinations 
of evolution parameters in split-step DTQW describes topological 
phases \cite{obuse2011topological, kitagawa2010exploring} 
and neutrino oscillation \cite{mallick2017neutrino, di2016quantum}. 
Indeed, coin parameters play a relevant role in the evolution 
of the state of the walker in the position space and, in turn, 
in controlling and engineering DTQWs. In this framework, a precise 
knowledge of the coin parameters is crucial information for 
quantum simulations and for further development in the use of 
quantum walks to model realistic quantum dynamics.   
\par
In the past, it has been determined that one coin parameter determines the group velocity of the walker's spread in position space  \cite{ahlbrecht2011asymptotic}. Therefore, by studying the standard deviation or group velocity of DTQW with one coin parameter, one can determine the value of parameter $\theta$ for unbounded DTQW. But the same does not hold for bounded or multi-parameter coin QW. Fisher information (FI) measures the amount of information that can be obtained about the unknown parameters in the system by performing measurement on the system, individually. Therefore, it can be used to obtain information of all the coin parameters when the coin operator is a general SU(2) operator with three parameters coin operation, multi-coin operation, or the bounded case. Here, we first develop a technique to calculate Fisher information in 
DTQW using a one parameter coin operator, and then extend to two-coin quantum 
walk. In this paper, we first consider a coin operator with one parameter 
$\theta$ and address the evolution of bounded and unbounded DTQWs in one dimension. 
Our aim is to design optimal estimation techniques for the coin 
parameter based on measurements performed on the sole position 
space of the walker. Our approach belongs to the class of 
protocols usually referred to as quantum probing\,\cite{smirne2013quantum, benedetti2014quantum, paris2014quantum, benedetti2014characterization, rossi2015entangled, tamascelli2016characterization, seveso2017can, bina2018continuous, benedetti2018quantum, troiani2018universal, beggi2018probing, pizio2018quantum}, which proved useful to 
precisely extract information upon exploiting the inherent 
sensitivity of quantum systems to external perturbations.
\par
We use the FI to 
quantify the information about a parameter $\theta$
which may be extracted by performing a given measurement
on a quantum system. In particular, we consider 
the FI $F_w (\theta)$ of the generic measurement 
performed on the walker's position degree of freedom. 
The maximum of $F_w (\theta)$ over all the
possible measurements is the so-called quantum Fisher 
information $H_w (\theta)$ (QFI), which 
quantifies the ultimate quantum bound to the extractable 
information, i.e., the overall information encoded onto 
the state of the system. We also evaluate the full QFI 
$H_f (\theta)$, i.e., the QFI of the position plus coin state, 
in order to assess the overall performances of 
measurements on the sole position space of the walker, compared
to measurements having access to the full quantum state.
Our results show that the walker QFI $H_w (\theta)$ 
increases as $t^2$, as it happens for the full QFI 
$H_f (\theta)$, meaning that the walker is a good 
probe for its coin operation parameter $\theta$. 
Additionally, the walker's position QFI 
$H_w (\theta)$ increases with $\theta$ and then 
decreases slowly up to $\pi/2$ ( 
and then mimics in the mirrored way, due to symmetry in 
the coin operator up to $\pi$). Finally, we analyze in 
some detail the performances of position measurement on 
the walker; i.e., we assess
how much information on the coin parameter may be extracted by looking
at the probability distribution 
of the walker at a given time. We also present QFI in position space for split-step quantum walk, where we have two coin parameters; we can see that QFI $H_w(\theta)$ helps us to estimate the coin parameters.
\par
The paper is structured as follows. In Sec. \ref{sec2}, we describe bounded and unbounded DTQWs and the evolution operators governing 
their dynamics. In Sec. \ref{sec3} we review quantum estimation 
theory, describe a method to numerically calculate the 
walker's quantum Fisher information in DTQWs, and illustrate 
the main results of our analysis.
Sec. \ref{sec4} closes the paper with some concluding remarks. 
\section{\label{sec2}Evolution in discrete-time quantum walk}
\begin{figure}[h!]
\includegraphics[width=0.9\columnwidth]{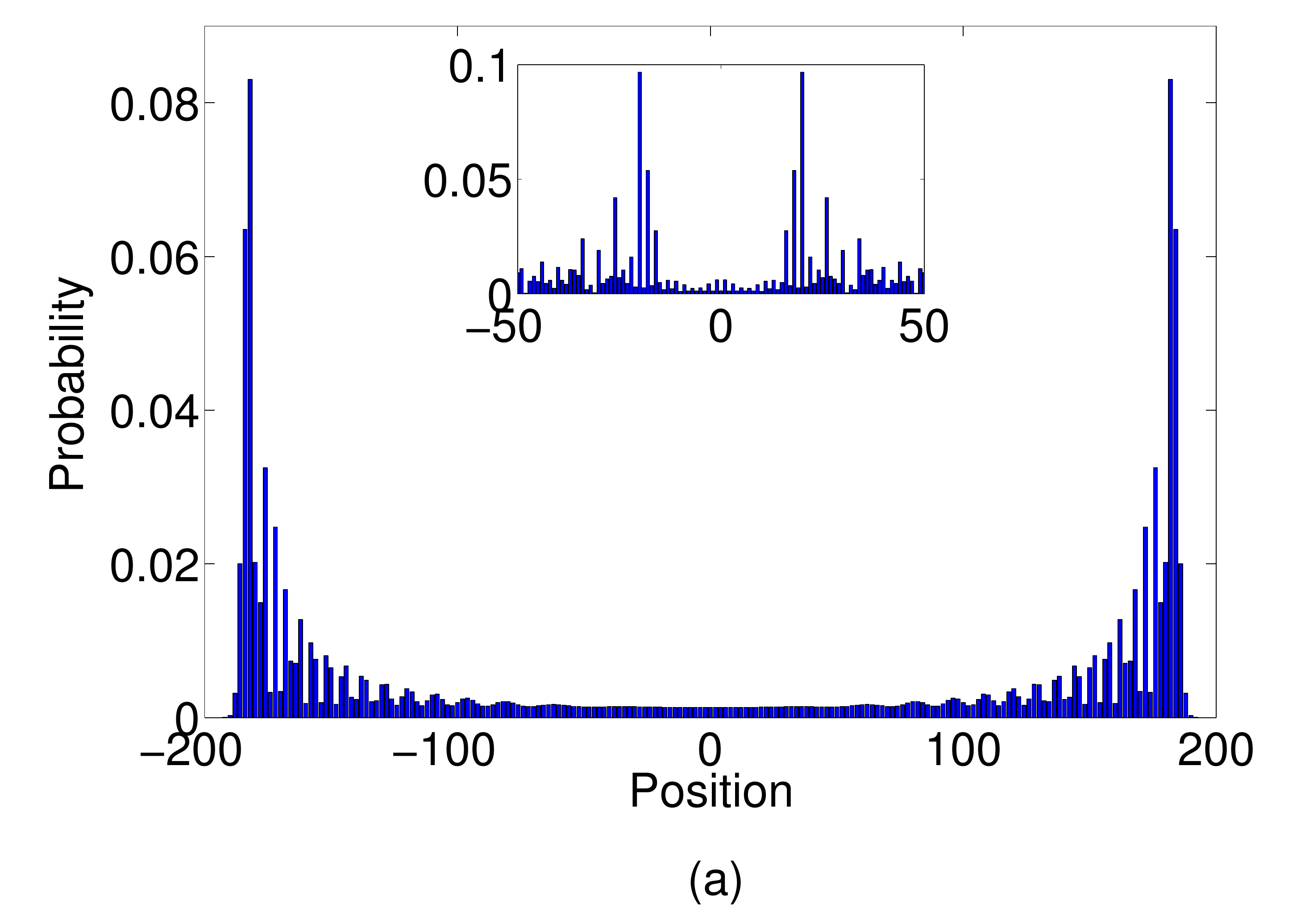}
\includegraphics[width=0.9\columnwidth]{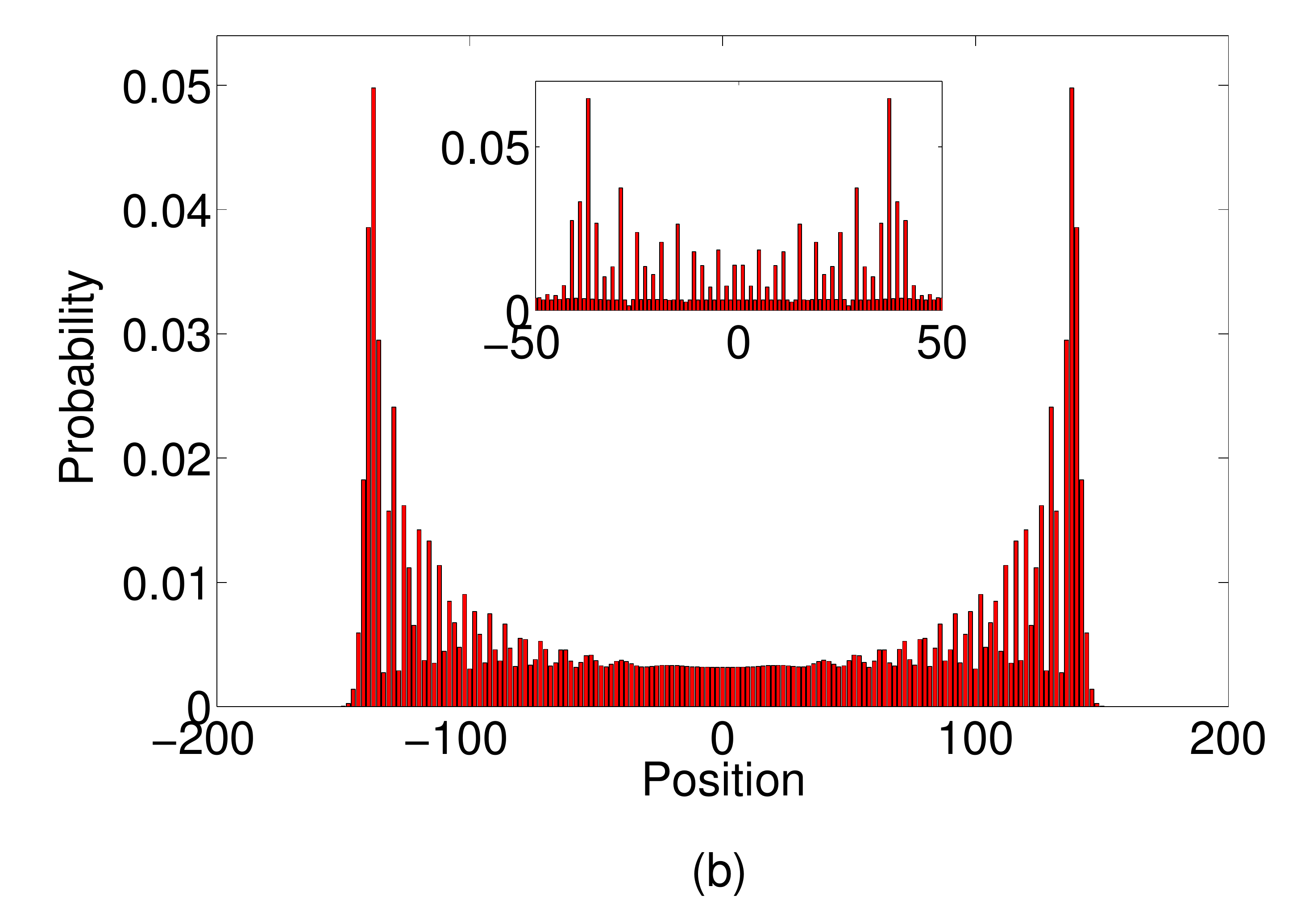}
\includegraphics[width=0.9\columnwidth]{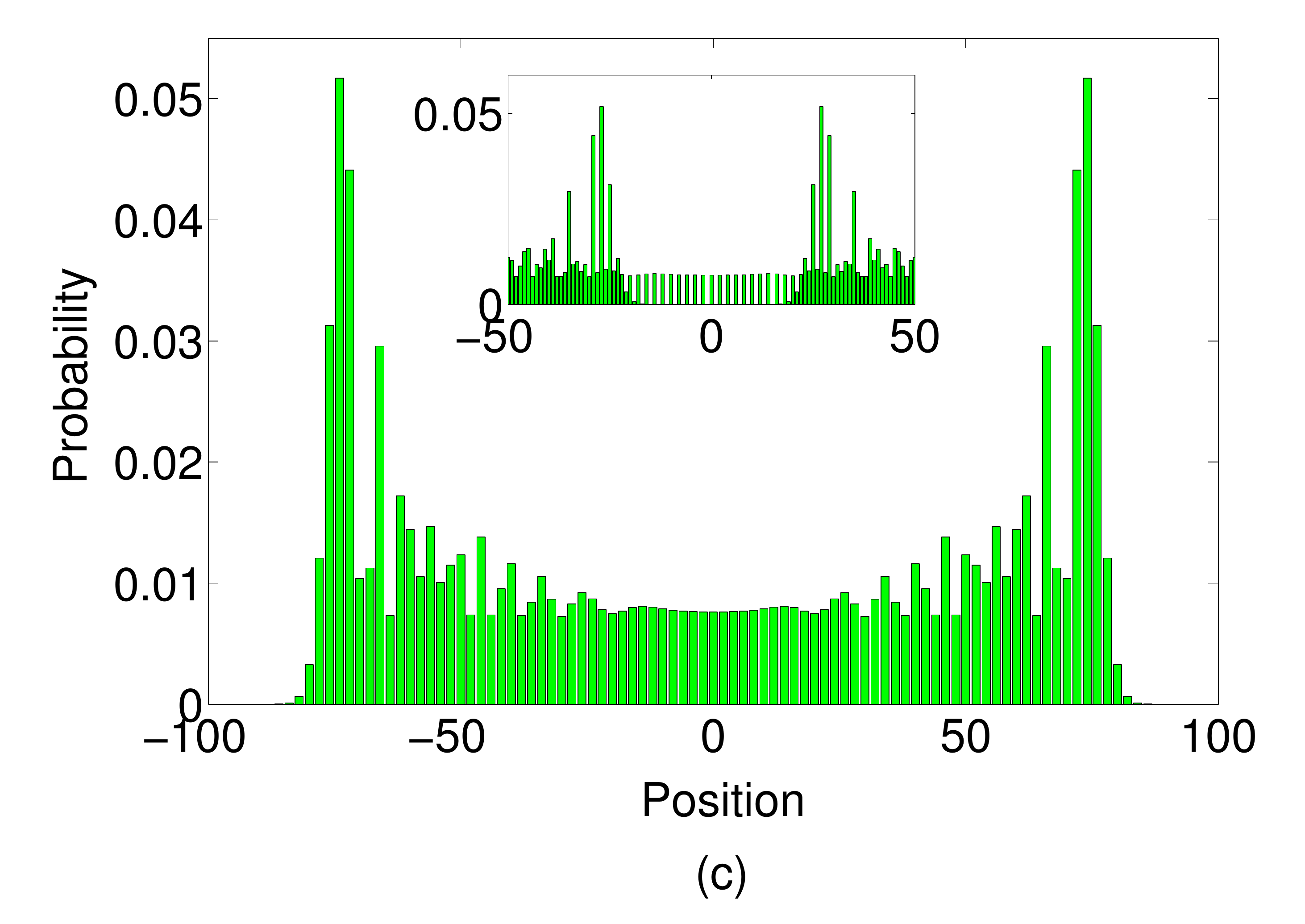}
\caption{Probability distribution of unbounded DTQW in one dimension
after 200 time steps for (a) $\theta = \pi/8$, (b) $\theta = \pi/4$ and (c) $\theta = 3\pi/8$, respectively.  The insets show the corresponding distributions for a DTQW bounded in the region 
$[-50,50]$. In both cases the initial state of the system 
has been set to $\frac1{\sqrt{2}}(\ket{\uparrow} + \ket{\downarrow}) 
\otimes \ket{x = 0} \equiv
|+\rangle\otimes |0\rangle$.}
\label{prob}
\end{figure}

DTQW of a single walker on a one dimensional lattice is 
defined on the Hilbert space $\mathcal{H} = \mathcal{H}_c 
\otimes \mathcal{H}_p$ where $\mathcal{H}_p$ and $\mathcal{H}_c$ 
are the position and the coin Hilbert spaces of the walker, respectively. The 
basis states of the coin Hilbert space are $\{ \ket{\uparrow}, \bra{\downarrow} \}$, which may be seen as the internal states of the walker. The position
 Hilbert space is spanned by the basis $\ket{x}$ where $x \in \mathbb{Z}$. The initial state of the
system is usually taken in the form
\begin{equation}
\ket{\Psi_{\hbox{\small in}}} = \alpha \ket{\uparrow} + \beta \ket{\downarrow}  
\otimes \ket{x = 0} ~~;
\quad |\alpha|^2 + |\beta|^2=1.
\end{equation}
Here $\alpha$ and $\beta$ are the amplitudes of the states 
$\ket{\uparrow}$ and $\ket{\downarrow}$, respectively. The 
evolution operator for discrete-time quantum walk is defined 
by the action of unitary quantum coin operation followed by 
a position shift operator. The single parameter coin operator 
is given by,
\begin{equation}\label{coinoperator}
 C_{\theta} = \begin{pmatrix}
 \cos\theta & -i \sin\theta \\
 -i \sin\theta  &  \cos\theta
 \end{pmatrix} \otimes \sum_{x} \ket{x}\bra{x}
\end{equation}
whereas the shift operator $S$ is defined with reference 
to the size of the region accessible by the walker.
{\em Unbounded} DTQWs are defined on a position Hilbert space 
of infinite size. The walker has no boundary condition on 
probability amplitude and the position shift operator is given by
\begin{align}
S_x = \sum_{x} \ket{\uparrow}\bra{\uparrow} \otimes \ket{x-1}\bra{x} + \ket{\downarrow}\ket{\downarrow} \otimes \ket{x+1}\bra{x}.
\end{align}
In Fig. \ref{prob} we show the probability distribution 
after 200 time steps for an unbounded DTQW using different values of 
coin parameter $\theta$. The smaller the value of $\theta$, the 
larger the spread of the probability distribution. 
{\em Bounded} DTQWs evolve instead on finite position Hilbert 
spaces, characterized by a finite number of sites and boundary 
conditions. In turn, the position shift operator is bounded between
 $[-a,a]$ with boundary condition $\ket{\Psi_{a+1}} = 
 \ket{\Psi_{-a-1}} = 0 $, where $a \in \mathbb{Z}$. In formula,
\begin{align}
S_x =& \ket{\downarrow}\bra{\uparrow} \otimes \ket{-a}\bra{-a} + \sum_{x= -a+1}^{a} \ket{\uparrow}\bra{\uparrow} \otimes \ket{x-1}\bra{x} \nonumber \\
+& \sum_{x= -a}^{a-1} \ket{\downarrow}\bra{\downarrow} \otimes \ket{x+1}\bra{x} + \ket{\uparrow}\bra{\downarrow} \otimes \ket{a}\bra{a}.
\end{align}
The insets of Fig.\,\ref{prob} show the probability distribution 
after 200 time step for a bounded DTQW and for different values of 
$\theta$. The position space is bounded between $-50$ and $50$. In this
case the shape of the probability distribution arises from the 
interplay of the coin operator and the bounded nature of the 
position space, and the spread cannot be simply characterized 
as a function of $\theta$, as it was for unbounded walk. 
\par
In general, after $t$ steps in the evolution, 
the overall state of the particle will be of the form,
\begin{align}
\ket{\Psi_t} = \Big(S_x\,C_\theta \Big)^t \ket{\Psi_{in}} = \sum_{x} \Big(\mathcal{A}_{x,t} \ket{\uparrow} + \mathcal{B}_{x,t} \ket{\downarrow} \Big) \otimes |x\rangle.
\end{align} 
where $\mathcal{A}_{x,t}$ and $\mathcal{B}_{x,t}$ are the  
amplitudes of the states $\ket{\uparrow}$ and $\ket{\downarrow}$ 
at position $x$ at time $t$, respectively. The $\mathcal{A, B}$ 
coefficients are in turn linked by the iterative relations
\begin{align}
\begin{pmatrix}
\mathcal{A}_{x,t} \\
\mathcal{B}_{x,t}
\end{pmatrix} &= \begin{pmatrix}
\cos\theta & -i\sin\theta \\
0 & 0
\end{pmatrix} \begin{pmatrix}
\mathcal{A}_{x+1,t-1} \\
\mathcal{B}_{x+1,t}
\end{pmatrix} \nonumber \\
&+ \begin{pmatrix}
0 & 0\\
-i\sin\theta & \cos\theta
\end{pmatrix} \begin{pmatrix}
\mathcal{A}_{x-1,t-1} \\
\mathcal{B}_{x-1,t-1}
\end{pmatrix}
\end{align}
for both, unbounded and bounded discrete-time quantum walk 
(when the walker is away from the boundary). Therefore, 
the probability of finding the particle at position $x$ 
and at time $t$ is given by,
\begin{align}
P(x,t) = |\mathcal{A}_{x,t}|^2 + |\mathcal{B}_{x,t}|^2.
\end{align} 
\section{\label{sec3} Quantum estimation in discrete-time quantum walk}
The Fisher information provides a measure of the amount 
of information that the observable $X$ carries about a 
parameter $\xi$, usually a quantity of interest, influencing
its probability distribution $p(x|\xi)$ \cite{paris2009quantum}. 
In more detail, the Fisher information $F(\xi)$ of a 
conditional distribution $p(x|\xi)$ is given by,
\begin{align} \label{FI}
F(\xi) = \int\! dx\, p(x|\xi)\, \Big[\frac{\partial 
\log p(x|\xi)}{\partial \xi} \Big]^2\,,
\end{align}
where, as mentioned above, $p(x|\xi)$ is the probability of 
obtaining the outcome $x$ from the measurement of $X$ when the 
true value of the parameter is $\xi$. 
If the available data for the observable $X$ are coming from 
$M$ repeated independent measurements of $X$, i.e., 
${\mathbf x}=(x_1,x_2,.....,x_M)$, then the overall probability 
of the sample (the likelihood) is $p({\mathbf x}|\xi) = 
\Pi_{k=1}^M p(x_k|\xi)$, which depends upon the parameter $\xi$ to be estimated. An {\em estimator} $\hat\xi ({\mathbf x})$ is a function
of the data sample, which provides an estimate of the value of the
parameter $\xi$. Since data fluctuate, the value of the estimator
fluctuates as well. The variance $\hbox{Var}_\xi \hat\xi$ of $\hat\xi$ provides a measure of the precision of the overall estimation procedure 
(i.e., the measurement of $X$ followed by the data processing $\hat\xi$).
The Cramer-Rao theorem states that the Fisher information poses a 
bound of the variance of $\hat\xi$
\begin{align}
\hbox{Var}_\xi \hat\xi \geq \frac{1}{M F(\xi)}\,. \label{cry}
\end{align}
The larger the value of $F(\xi)$ the greater the amount of information about $\xi$ 
that {\em may} be, {\em in principle} 
extracted from the measurement of $X$. The actual information on 
$\xi$ obtained from measuring $X$ instead 
depends on the estimator. An estimator saturating the Cramer-Rao 
bound of Eq. (\ref{cry})  is said to be {\em efficient}. In the 
following, we assume that an efficient estimator is available 
and compare the performances of different measurements in terms 
of their Fisher information.
\par
Let us now move to quantum measurements: According to Born's 
rule the conditional distribution $p(x|\xi)$ may be written as 
$p(x|\xi) = \hbox{Tr}[\Pi_x\,\rho_{\xi}]$ where, $\Pi_x$ is the 
probability operator-valued measure of the measured
quantity $X$, and the dependence on $\xi$ is encoded onto the 
preparation of the system undergoing the measurement, i.e., the
density  $\rho_{\xi}$.  An upper bound on the Fisher information 
of {\em any} quantum measurement may be obtained by introducing 
the symmetric logarithmic derivative (SLD) $L_{\xi}$, which 
satisfies the relation
\begin{align}\label{sld}
\frac12 \left (L_{\xi} \rho_{\xi} + \rho_{\xi} L_{\xi}\right) = \frac{\partial \rho_\xi}{\partial\xi}\,.
\end{align}  
Then, since $\partial_{\xi} p(x|\xi) = \hbox{Tr}[\partial_{\xi} \rho_{\xi} \Pi_{x}] = \hbox{Re}\left(Tr[\rho_{\xi} \Pi_{x} L_{\xi}]\right)$, 
the Fisher information may be rewritten in terms of $L_\xi$
and an upper bound on Fisher information, usually referred to 
as quantum Fisher information, may be found
\begin{align} \label{QFI}
F(\xi) \leq H(\xi) \equiv \hbox{Tr}[\rho_{\xi}\,L_{\xi}^2]
\end{align}
%
%
where $L_{\xi}$ is given in Eq. (\ref{sld}). For a pure state, $\rho_{\xi}^2 = \rho_{\xi}$ and therefore $\partial_{\xi} \rho_{\xi} = (\partial_{\xi} \rho_{\xi}) \rho_{\xi} + \rho_{\xi} (\partial_{\xi} \rho_{\xi})$ implies, $L_{\xi} = 2\partial_{\xi} \rho_{\xi}$. Hence, encoding $\rho_\xi = |\psi_\xi\rangle\langle\psi_\xi|$, the SLD reduces to $L_{\xi} = 2\partial_{\xi} \rho_{\xi}$. 



\subsection{\label{sec31}The full QFI $H_f(\theta)$ 
in discrete-time quantum walk}
\begin{figure}[h!]
\includegraphics[width=0.9\columnwidth]{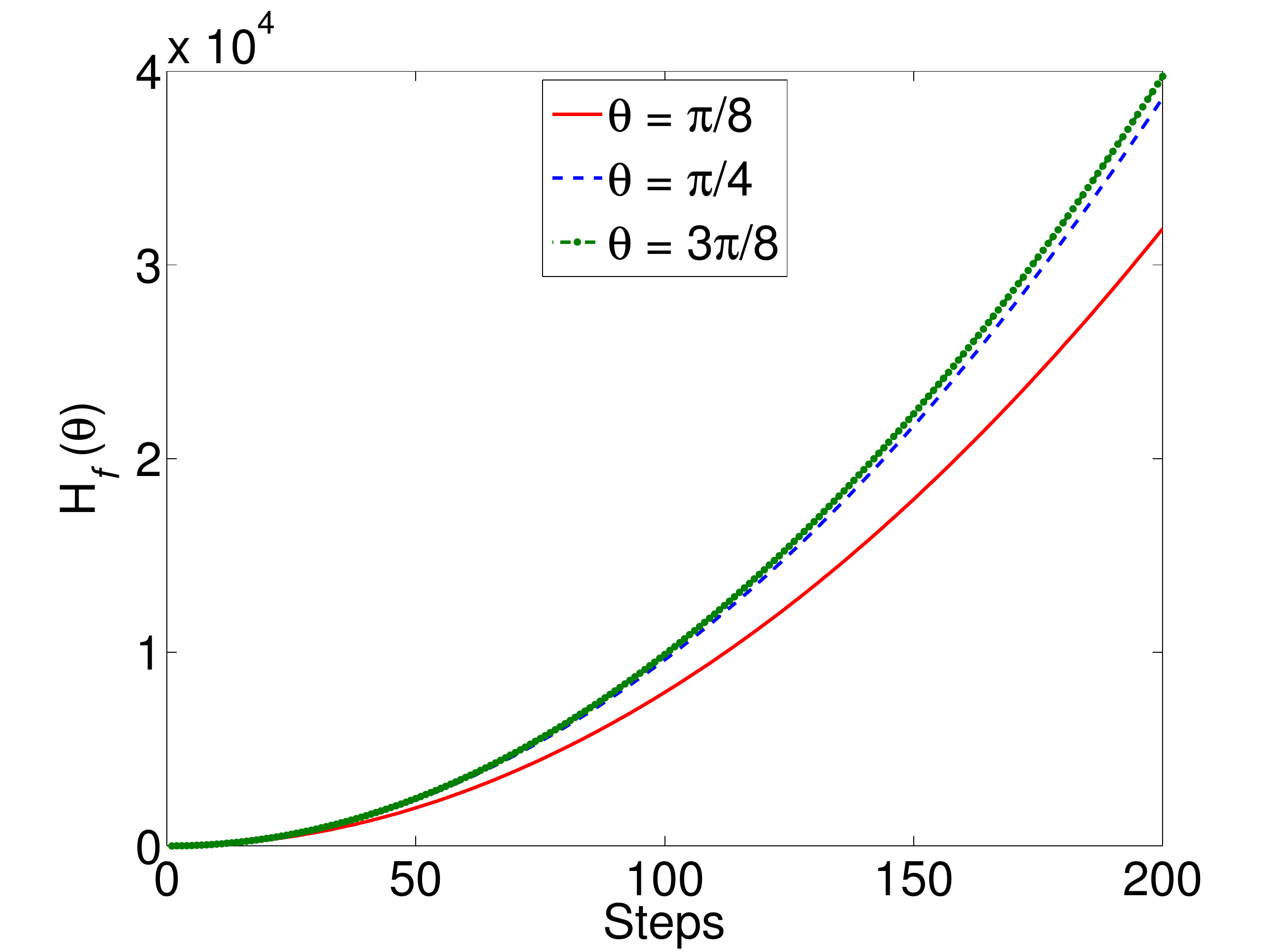}
\caption{The full QFI $H_f(\theta)$ as a function of time
for different values of $\theta$. The unit of $H_f(\theta)$ is inverse square of unit of $\theta$ therefore here it is $(rad)
^{-2}$ and the unit of $\theta$ is $radian$. The initial state of the 
system is $|+\rangle\otimes|0\rangle$. The full QFI 
$H_f(\theta)$ is the same for bounded and unbounded DTQWs. }
\label{QFIComplete}
\end{figure}

The density matrix of the full (coin plus position) state
in the complete Hilbert space $\mathcal{H} = \mathcal{H}_c 
\otimes \mathcal{H}_w$ at time $t$ is given by,
\begin{align} \label{eqrho}
\rho_{\theta} =  \ket{\Psi_{\theta}}\bra{\Psi_{\theta}} \equiv 
\begin{pmatrix}
\ket{\psi_{\theta}^{\uparrow}} \\
\ket{\psi_{\theta}^{\downarrow}}
\end{pmatrix} 
\begin{pmatrix}
\bra{\psi_{\theta}^{\uparrow}} \\
\bra{\psi_{\theta}^{\downarrow}}
\end{pmatrix}^T
\end{align}
where the size of the vector $\ket{\psi_{\theta}^{\uparrow}}$ 
and $\ket{\psi_{\theta}^{\downarrow}}$ is equal to the dimension
of the walker's position Hilbert space $\mathcal{H}_w$ and the 
dimension of $\rho_{\theta}$ is $2N$ where $N$ is the 
dimension of $\mathcal{H}_w$. This implies that 
$\partial_{\theta} \rho_{\theta}$ may be written as
\begin{align} \label{eqdrho}
\partial_{\theta} \rho_{\theta} &= \ket{\partial_{\theta} \Psi_{\theta}}\bra{\Psi_{\theta}} + \ket{\Psi_{\theta}}\bra{\partial_{\theta} \Psi_{\theta}} \nonumber \\
&= 
\begin{pmatrix}
\ket{\partial_{\theta} \psi_{\theta}^{\uparrow}} \\
\ket{\partial_{\theta} \psi_{\theta}^{\downarrow}}
\end{pmatrix} 
\begin{pmatrix}
\bra{\psi_{\theta}^{\uparrow}} \\ \bra{\psi_{\theta}^{\downarrow}}
\end{pmatrix}^T + 
\begin{pmatrix}
\ket{\psi_{\theta}^{\uparrow}} \\
\ket{\psi_{\theta}^{\downarrow}}
\end{pmatrix} 
\begin{pmatrix}
\bra{\partial_{\theta} \psi_{\theta}^{\uparrow}} 
\\ \bra{\partial_{\theta} \psi_{\theta}^{\downarrow}}
\end{pmatrix}^T
\end{align}
and $\ket{\partial_{\theta} \Psi_{\theta}}$ at time $t$ is given by,
\begin{align}
\ket{\partial_{\theta} \Psi_{\theta}(t)} &= S_x\,C_\theta \ket{\partial_{\theta} \Psi_{\theta}(t-1)} 
+ S_x(\partial_{\theta}C_\theta) \ket{\Psi_{\theta}(t-1)}
\end{align}  
where,
\begin{align}
\partial_{\theta} C_{\theta} = \begin{pmatrix}
-\sin\theta & -i \cos\theta \\
-i \cos\theta & -\sin\theta
\end{pmatrix} \otimes \sum_{x} \ket{x}\bra{x}.
\end{align}
As a consequence, if at a given time $t$, 
we have the amplitude $\Psi_x = (\psi_{x}^{\uparrow}; 
\psi_{x}^{\downarrow}) = (\mathcal{A}_{x}, 
\mathcal{B}_{x})$, then the iterative form 
for $\ket{\partial_{\theta} \Psi_{\theta}(t)}$ is 
given by,
\begin{align}
\begin{pmatrix}
\partial_{\theta} \mathcal{A}_{t,x} \\
\partial_{\theta} \mathcal{B}_{t,x}
\end{pmatrix} 
=&  
\begin{pmatrix}
\cos(\theta) & -i\sin(\theta)\\
0 & 0
\end{pmatrix} 
\begin{pmatrix}
\partial_{\theta} \mathcal{A}_{t-1,x+1} \\
\partial_{\theta} \mathcal{B}_{t-1,x+1}
\end{pmatrix} 
\nonumber \\
&+
\begin{pmatrix}
0 & 0 \\
-i\sin\theta & \cos\theta
\end{pmatrix} \begin{pmatrix}
\partial_{\theta} \mathcal{A}_{t-1,x-1} \\
\partial_{\theta} \mathcal{B}_{t-1,x-1}
\end{pmatrix} 
\nonumber \\
&+
\begin{pmatrix}
-\sin\theta & -i\cos\theta\\
0 & 0
\end{pmatrix} \begin{pmatrix}
 \mathcal{A}_{t-1,x+1} \\
 \mathcal{B}_{t-1,x+1}
\end{pmatrix} 
\nonumber \\
&+ 
\begin{pmatrix}
0 & 0 \\
-i\cos\theta & -\sin\theta
\end{pmatrix} \begin{pmatrix}
 \mathcal{A}_{t-1,x-1} \\
 \mathcal{B}_{t-1,x-1}
\end{pmatrix}\,.
\end{align} 
Upon substituting Eqs.\,\eqref{eqrho} and \,\eqref{eqdrho} in Eq.\,\eqref{QFI} we obtain the quantum Fisher information $H_f(\theta)$
in the complete Hilbert space $\mathcal{H} = \mathcal{H}_c \otimes \mathcal{H}_w$, i.e., the information extractable from the full 
quantum state of the walker's position plus coin system. In Fig.\,\ref{QFIComplete} 
we show $H_f(\theta)$ for unbounded and bounded DTQWs 
after 200 time steps. The full QFI $H_f(\theta)$ increases as 
$t^2$ with time and it is the same for bounded and unbounded 
DTQWs.
\subsection{\label{sec32} The walker's position space QFI $H_w(\theta)$ 
in discrete-time quantum walk}
The density matrix of the sole position space of the walker is obtained by 
tracing out the coin degree of freedom from Eq.\,
\eqref{eqrho}. We have
\begin{align}\label{rhow}
\rho_w(\theta) = \ket{\psi^{\uparrow}_{\theta}}\bra{\psi^{\uparrow}_{\theta}} + \ket{\psi^{\downarrow}_{\theta}}\bra{\psi^{\downarrow}_{\theta}}\,,
\end{align}
and, in turn,
\begin{align}
\partial_{\theta} \rho_{w}(\theta) &= \ket{\partial_{\theta} \psi_{\theta}^{\uparrow}}\bra{\psi_{\theta}^{\uparrow}} + \ket{\psi_{\theta}^{\uparrow}}\bra{\partial_{\theta} \psi_{\theta}^{\uparrow}} \nonumber\\
&+ \ket{\partial_{\theta} \psi_{\theta}^{\downarrow}}\bra{\psi_{\theta}^{\downarrow}} + \ket{\psi_{\theta}^{\downarrow}}\bra{\partial_{\theta} \psi_{\theta}^{\downarrow}}
\end{align}
which is equal to tracing out the coin from the derivative of the 
full density matrix in complete Hilbert space; i.e., tracing out 
the coin from Eq.\,\eqref{eqdrho},
\begin{align} \
\partial_{\theta} \rho_{w}(\theta) = \hbox{Tr}_c \Big[
\partial_{\theta} \rho_{\theta} \Big]\,.
\end{align}
%
 The density matrix in position space will be in a mixed state. In the mixed state, $\rho_{w}^2(\theta) = \rho_{w}(\theta) + \rho_1(\theta)$, where $\rho_1(\theta) = \frac{\epsilon}{2}\int d\theta (\lambda \rho_{w}(\theta) + \rho_{w}(\theta) \lambda) + O(\epsilon^2)  $. Here $\epsilon$ represents the fluctuation in the measure of how mixed the density matrix in position is and $\lambda$ can be calculated by taking the partial derivative of $(\rho_{w}^2(\theta) - \rho_w(\theta))$ with respect to $\theta$ when the fluctuation is very small. The value of $\lambda$ is $(\partial_{\theta} \rho_w(\theta) - 1/2)$ when $(\partial_{\theta}\rho_w(\theta) - \rho_w(\theta)) \rightarrow 0$. Therefore  the SLD is $L_{\theta} = 2\partial_{\theta} \rho_{w}(\theta)  + \epsilon \lambda $ 
. This implies that,
\begin{align}
L^2 &= 4 (\partial_{w}(\theta) \rho_{w}(\theta))^2 + 2 \epsilon \Big[ \lambda \partial_{\theta} \rho_{w}(\theta) + (\partial_{\theta} \rho_{w}(\theta)) \lambda \Big] + O(\epsilon^2) \nonumber \\
L^2 &\approx 4(\partial_{\theta} \rho_{w}(\theta))^2 + 2\Big[ (L - 2\partial_{\theta} \rho_{w}(\theta)) \partial_{\theta} \rho_{w}(\theta) \nonumber \\
&+ \partial_{\theta} \rho_{w}(\theta) (L - 2\partial_{\theta} \rho_{w}(\theta))  \Big] \nonumber \\
L^2 &\approx -4(\partial_{\theta} \rho_{w}(\theta))^2 + 2(\partial_{\theta} \rho_{w}(\theta) L + L \partial_{\theta} \rho_{w}(\theta))
\end{align}
and therefore quantum Fisher information in the mixed state can be given by,
\begin{align}
H_{w} &= \hbox{Tr}[\rho_{w}(\theta) L^2] \nonumber \\
&\approx -4\hbox{Tr}[\rho_{w}(\theta) (\partial_{\theta} \rho_{w}(\theta))^2] + 2\hbox{Tr}[\rho_{w}(\theta)(\partial_{\theta} \rho_{w}(\theta) L \nonumber \\
&+ L \partial_{\theta} \rho_{w}(\theta))] \nonumber \\
&= 2\hbox{Tr}[\partial_{\theta}\rho_{w}(\theta)( L \rho_{w}(\theta) + \rho_{w}(\theta) L)] \nonumber \\
&- 4\hbox{Tr}[\rho_{w}(\theta) (\partial_{\theta} \rho_{w}(\theta))^2] \nonumber \\
&= 4\hbox{Tr}\Big[ (\partial_{\theta} \rho_{w}(\theta))^2 (\mathbb{I} - \rho_{w}(\theta)) \Big]
\end{align}
This expression for quantum Fisher information for the mixed state is obtained with an approximation that the higher powers of $\epsilon$ are very small and thus ignoring them. 

\begin{figure}[h!]
\includegraphics[width=0.49\columnwidth]{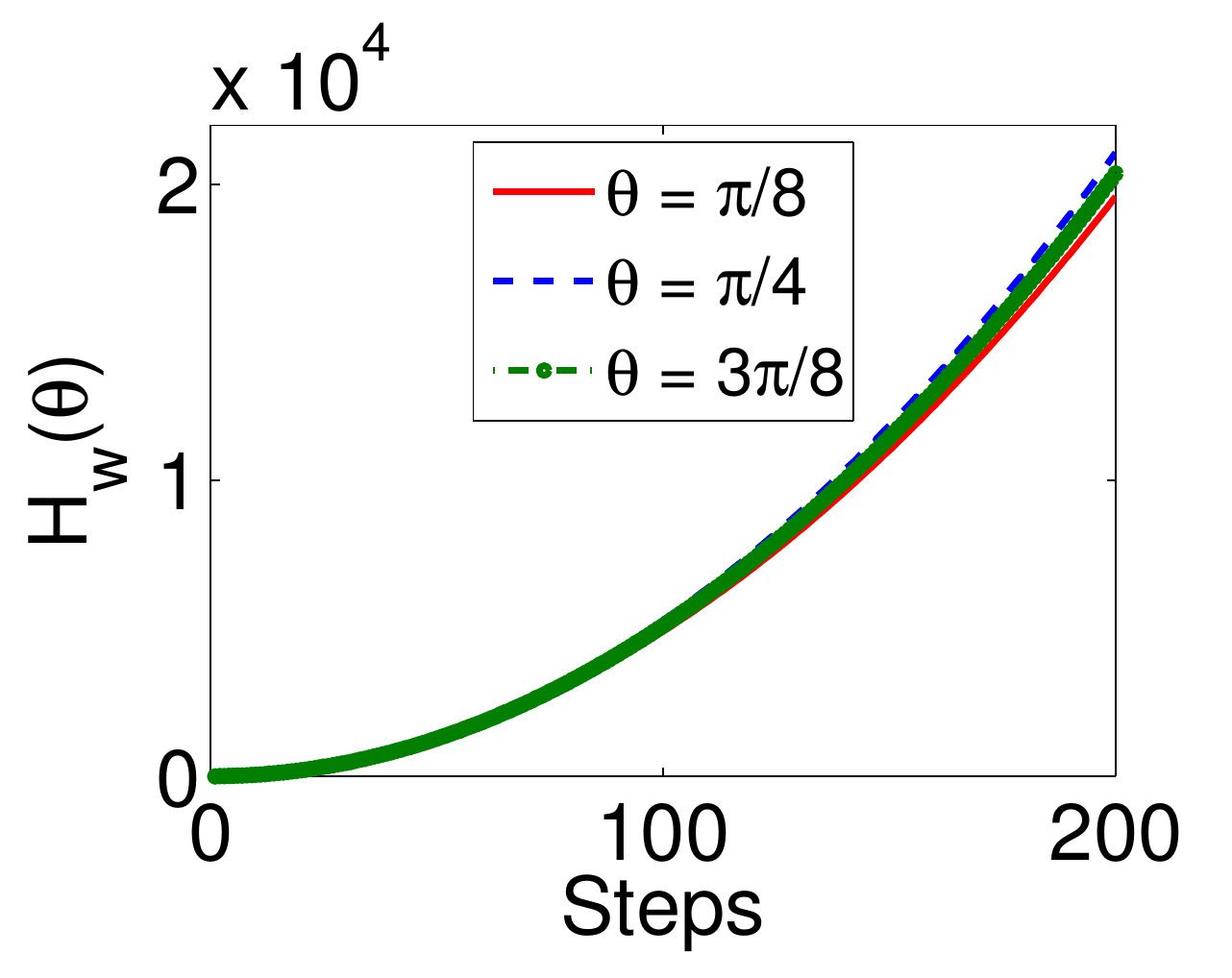} \label{fig:QFIUBP}
\includegraphics[width=0.49\columnwidth]{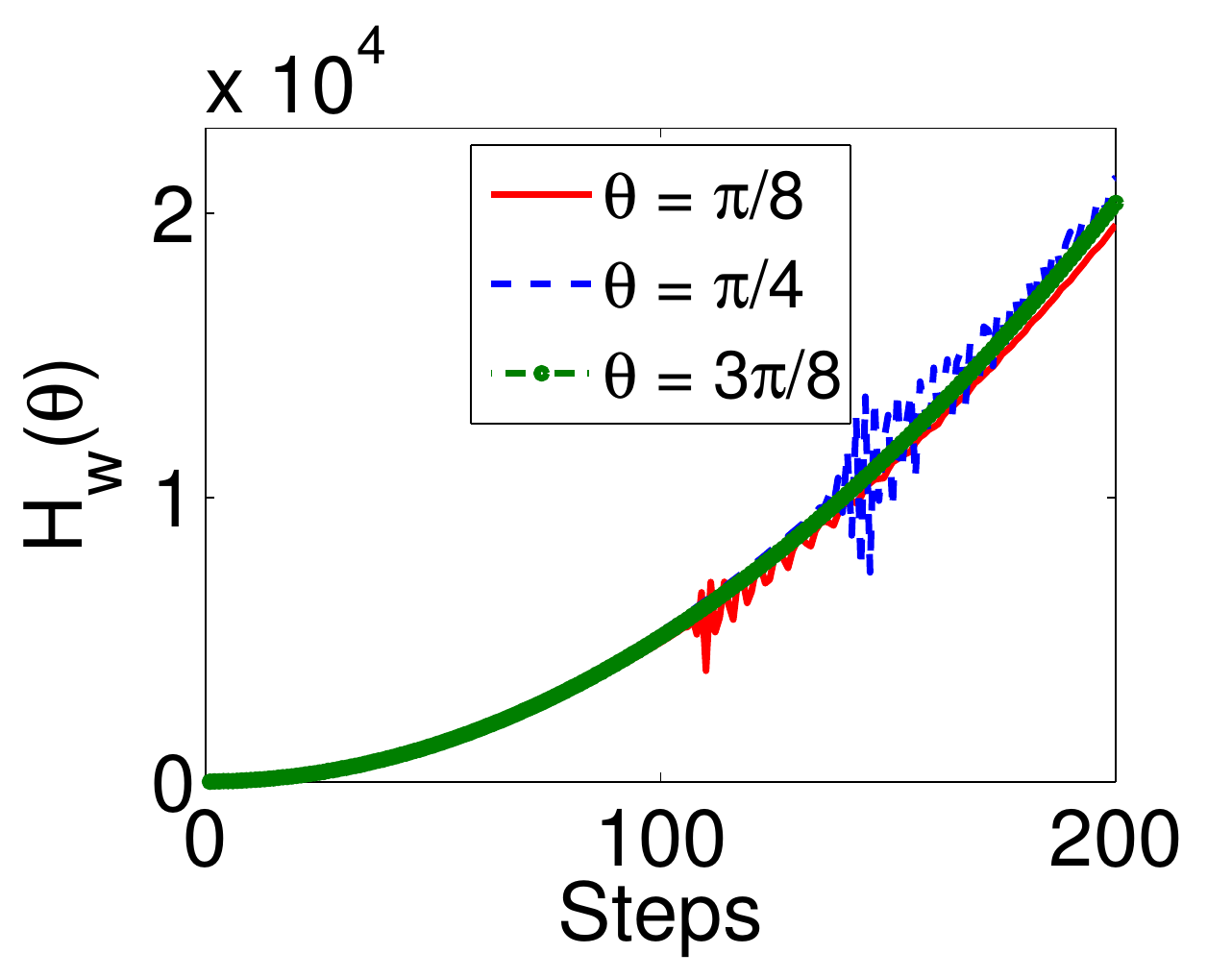} \label{fig:QFIBP}
\caption{The walker's position space QFI $H_w(\theta)$ as a function of time
for different values of $\theta$. The unit of $H_w(\theta)$ is inverse square of unit of $\theta$ therefore here it is $(rad)^{-2}$ and unit of $\theta$ is $radian$. Due to interference effects, 
we see clear differences between the walker's position space QFIs of bounded and unbounded 
DTQWs. The initial state of the system in both cases is $|+\rangle 
\otimes |0\rangle$.}
\label{fig:QFIP}
\end{figure}
\par
\begin{figure}[h!]
\includegraphics[width=0.49\columnwidth]{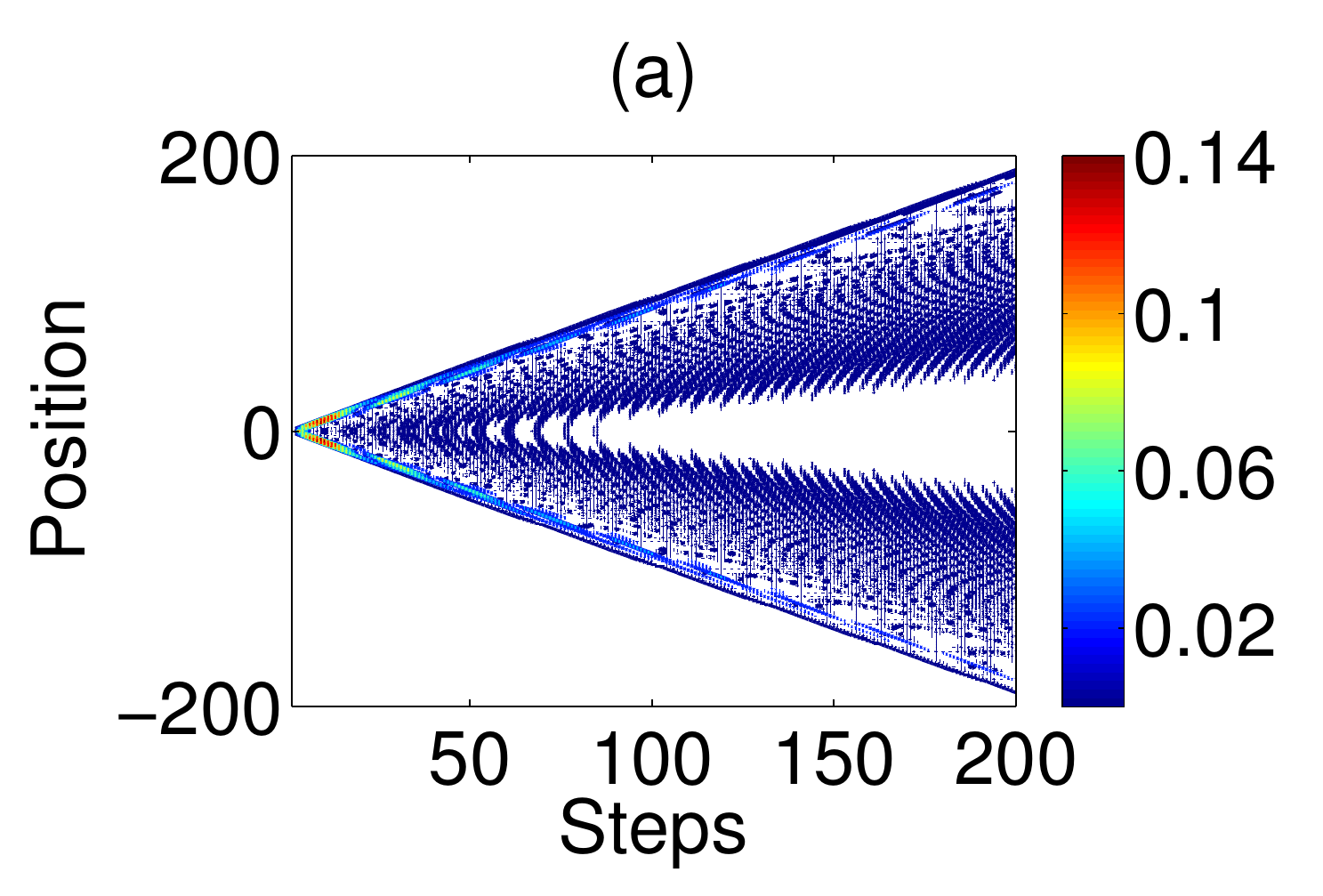} \label{IntUBpi8}
\includegraphics[width=0.49\columnwidth]{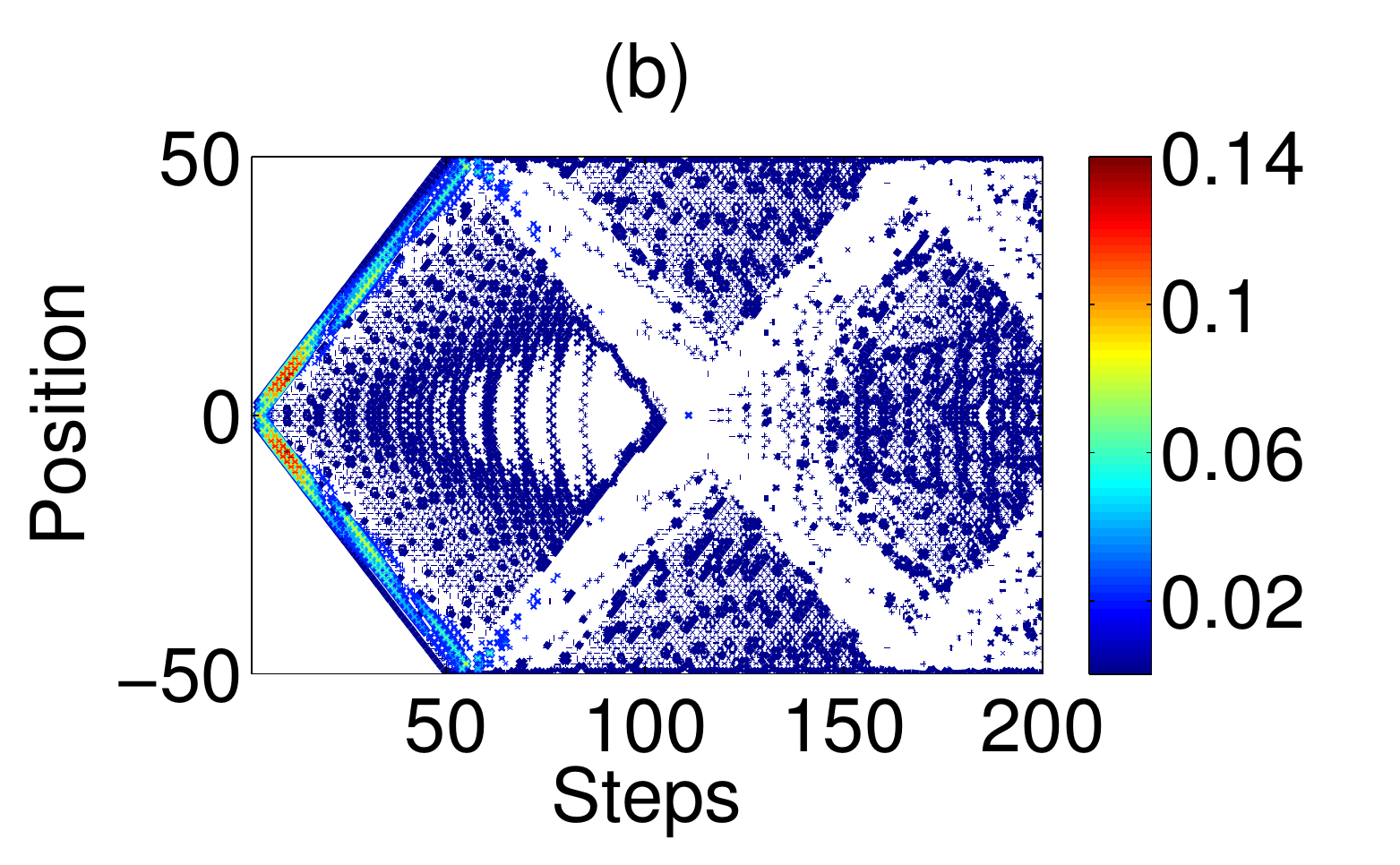} 
\label{IntBpi8}
\includegraphics[width=0.49\columnwidth]{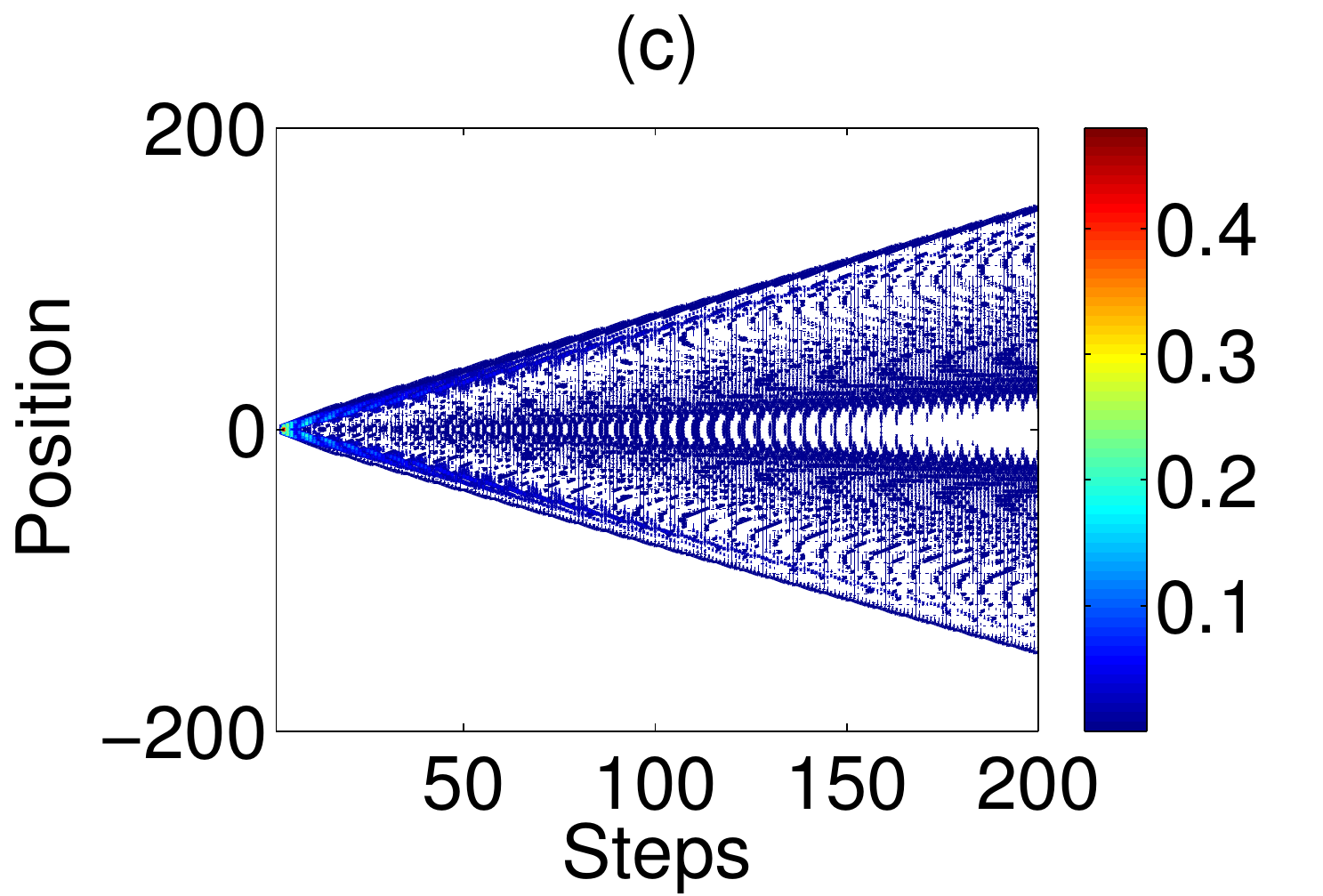} \label{IntUBpi4}
\includegraphics[width=0.49\columnwidth]{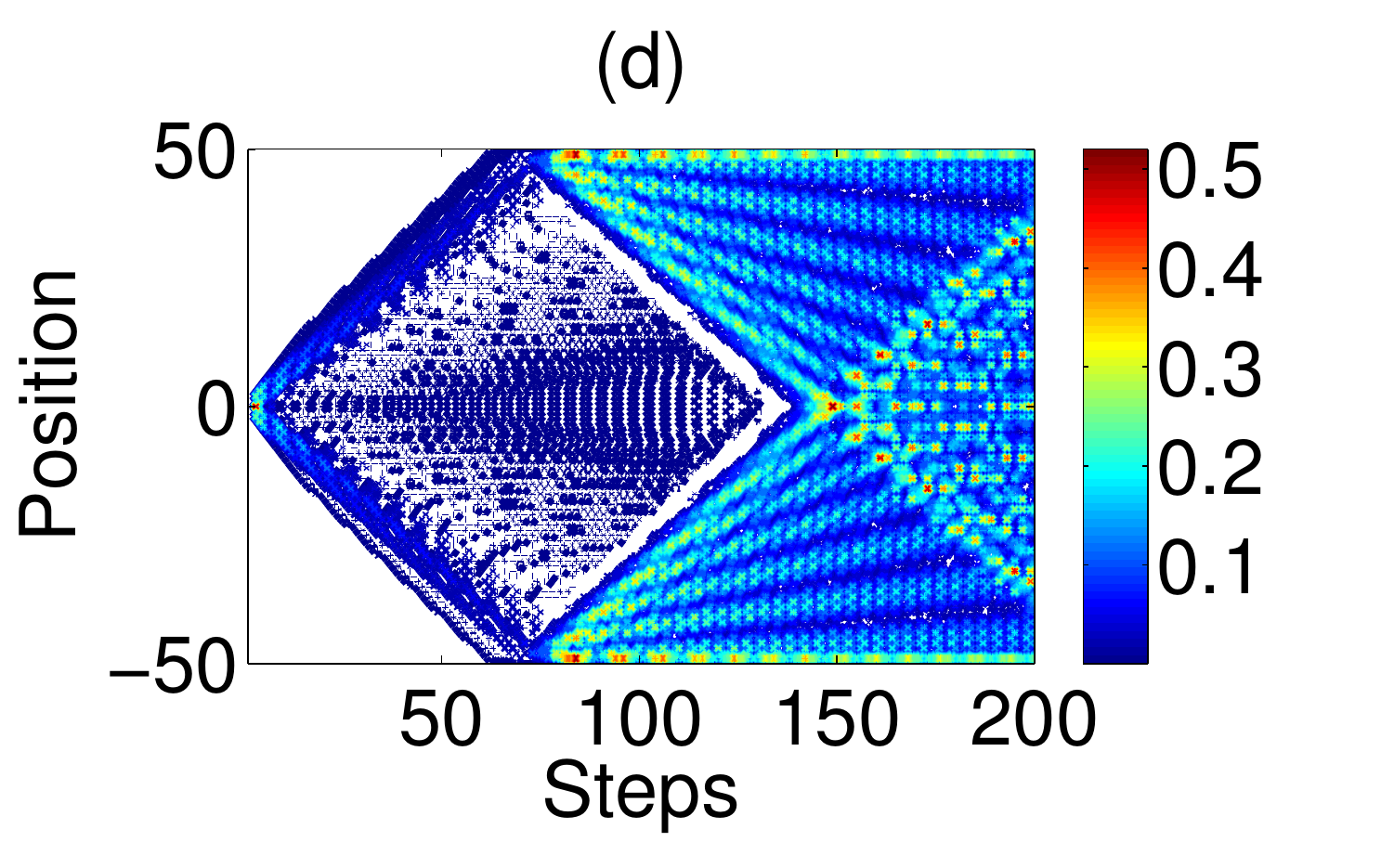} 
\label{IntBpi4}
\includegraphics[width=0.49\columnwidth]{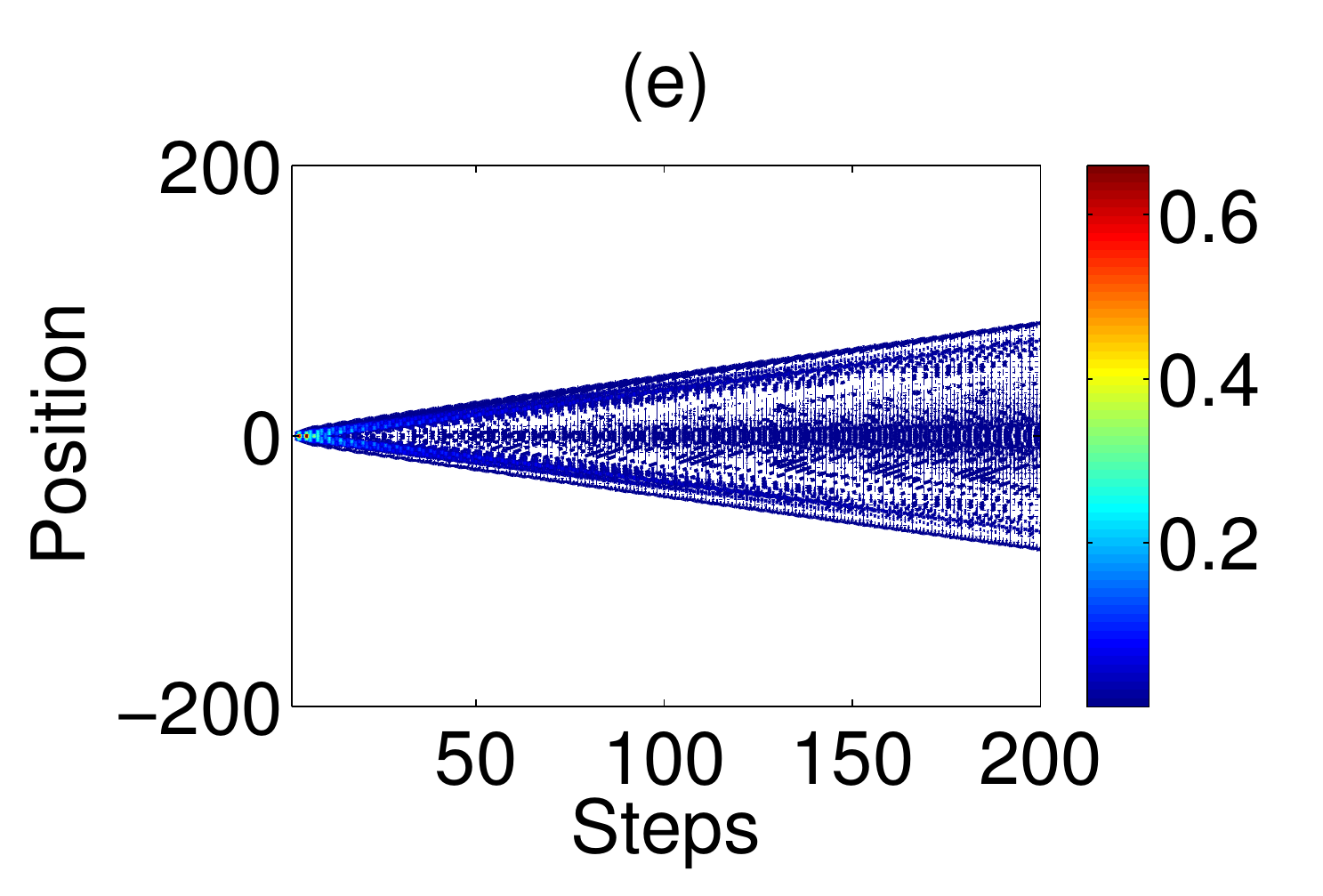} \label{IntUB3pi8}
\includegraphics[width=0.49\columnwidth]{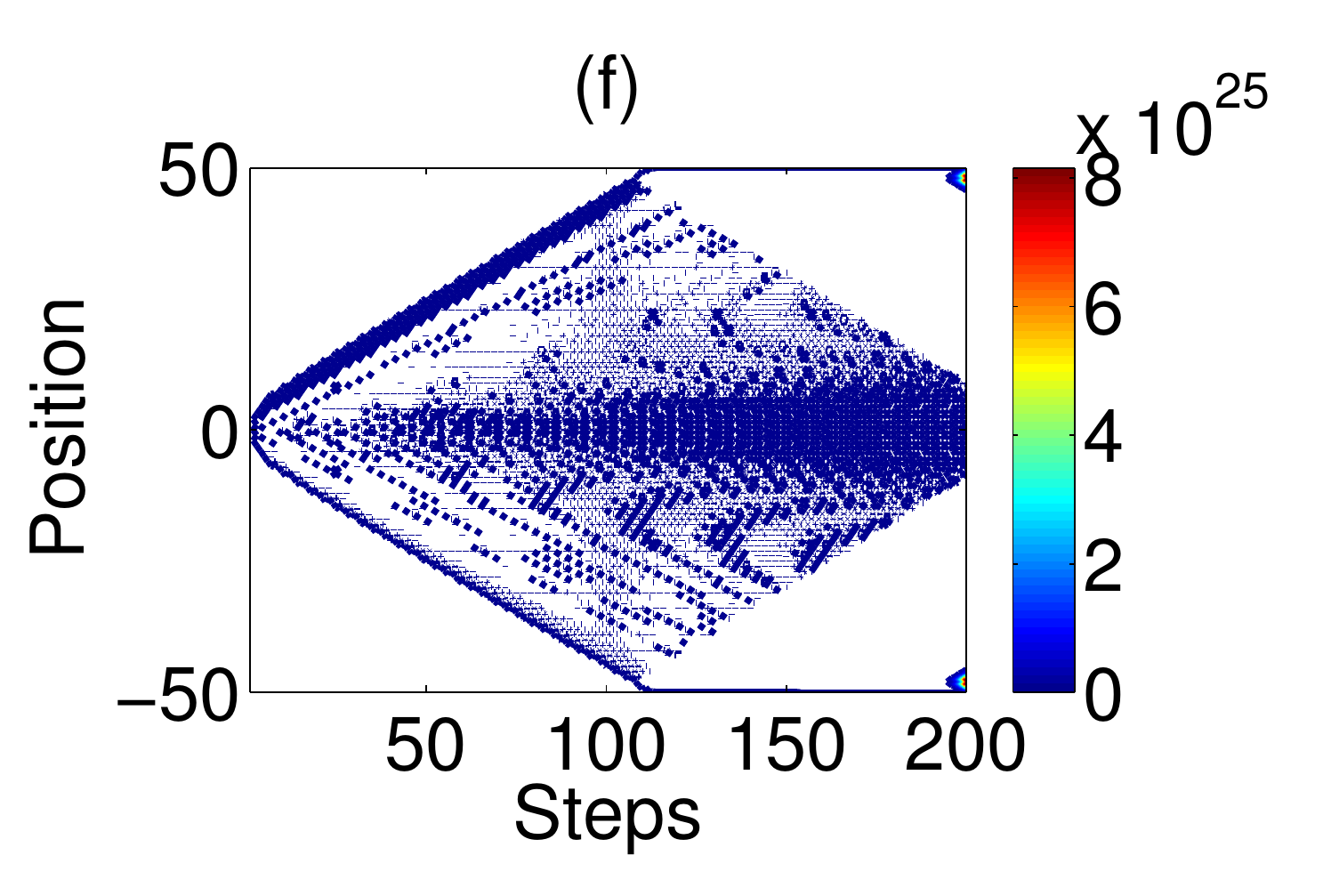} 
\label{IntB3pi8}
\caption{Density plot of the degree of interference $\mu$ for bounded and 
unbounded DTQWs as a function of the time steps and the position. Panels 
(a) and (b) describe the behavior of $\mu$ for $\theta = \pi/8$; left panel
for unbounded and right panel for bounded DTQW, respectively. Similarly, 
(c) and (d) correspond to the dynamics for $\theta = \pi/4$  and (e) 
and (f) correspond to $\theta = 3\pi/8$. The bounded walker is moving in 
the interval $[-50,50]$. The initial state of the walker is $\ket{+} 
\otimes \ket{x=0}$.}
\label{Int}
\end{figure}

Figure \,\ref{fig:QFIP} illustrates the behavior of $H_w(\theta)$ as a function
of time for different values of $\theta$ and for both bounded and unbounded DTQWs.
As we have seen for the full QFI $H_f(\theta)$, also $H_w(\theta)$ increases as 
$t^2$. For $t$ large enough (say, $t> 10$), we have $H_w(\theta) = \kappa t^2$, 
with the constant depending only on $\theta$, $\kappa \equiv \kappa(\theta)$. 
However, some striking differences between the two cases appear after 
$2a$ time steps, $[-a,a]$ being the spatial interval for bounded DTQW. Those 
differences may be traced back to interference \cite{singh2017interference} 
and recurrence \cite{vstefavnak2008recurrence, chandrashekar2010fractional} 
in the position space. In order to illustrate this phenomenon, in Fig.\,\ref{Int} 
we show the time evolution of the so-called {\it degree of interference} in 
the position Hilbert space, i.e the quantity 
\begin{align}
\mu_{x,t+1} & = \Bigg|\sin\theta \cos\theta\,  \Big[\rho^{\uparrow \downarrow}_{t}(x+1,x+1) \nonumber \\ &\qquad\qquad\qquad  
- \rho^{\uparrow \downarrow}_{t}(x-1,x-1)\Big] \nonumber \\
& + \sin\theta \cos\theta  \Big[\rho^{\downarrow \uparrow}_{t}(x+1,x+1) 
\nonumber \\ &\qquad\qquad\qquad   - \rho^{\downarrow \uparrow}_{t}(x-1,x-1)\Big]\Bigg|\,,
\end{align} 
defined for any site $x$ at the time $t+1$.
As it is apparent by comparing Figs. \ref{fig:QFIP} and \ref{Int}, the 
difference between $H_w(\theta)$ of bounded and unbounded DTQWs starts 
to appear in correspondence of the time step for which also the degrees 
of interference of the two cases start to differ, since the interference 
at a position $x$ at time $t$ in bounded walk is not only due to the 
neighboring sites but also due to the multiple sites. For example, 
in Fig.\,\ref{Int}-(b), it can be seen that for $\theta = \pi/8$ 
the degree of interference initially spreads over the position 
space with time, and then starts to come back at the initial position 
state. After $t=100$ time steps, interference between the reflected 
waves dominates, as we have seen for the QFI $H_w(\theta)$. A similar
behavior (see  Fig.\,\ref{fig:QFIP}  and the other panels of Fig.\,\ref{Int}) may 
be observed for 
the other values of $\theta$.

\par
\par
\begin{figure}[h!]
\includegraphics[width=0.9\columnwidth]{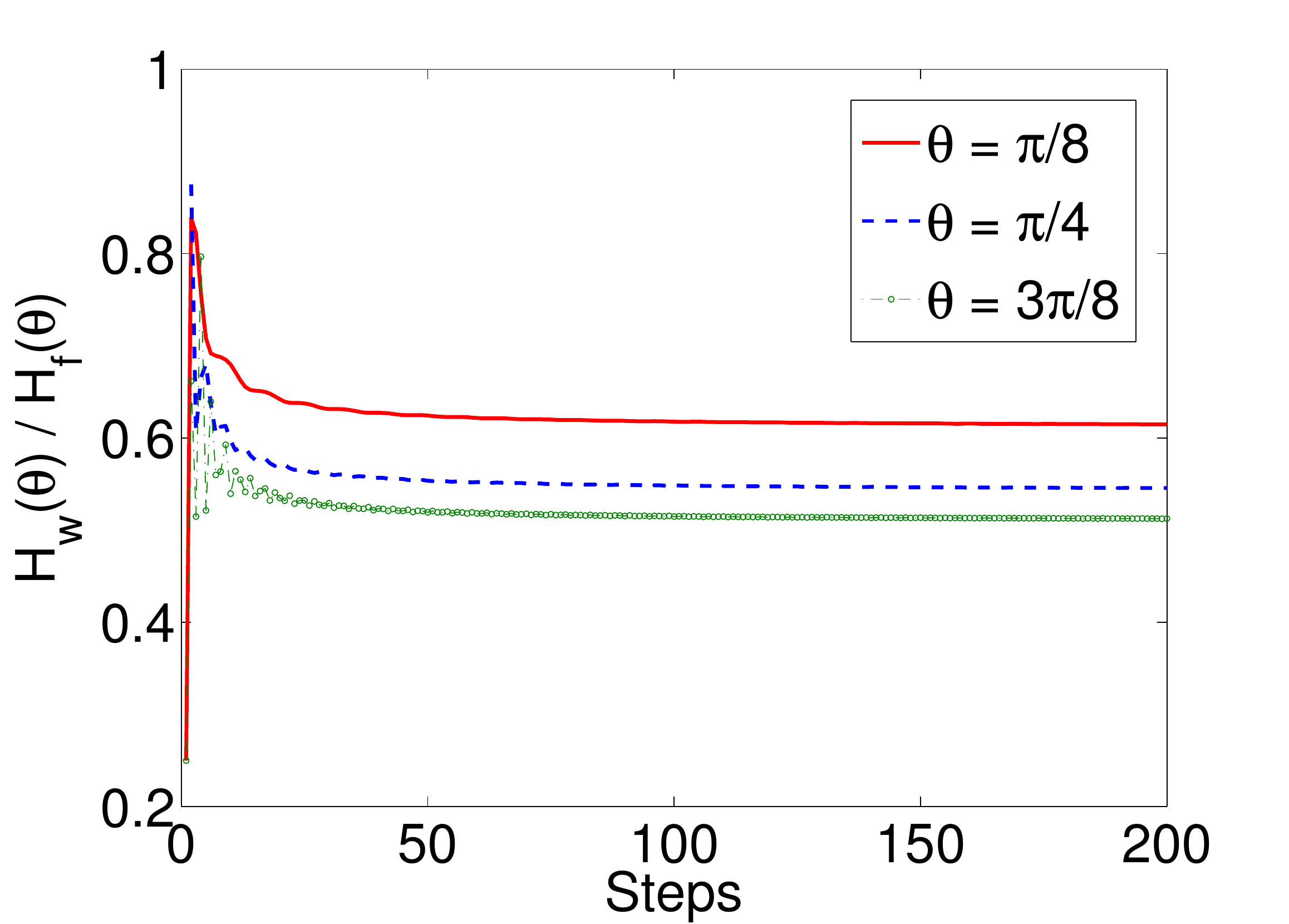}
\caption{Ratio of quantum Fisher information in position 
Hilbert space $\mathcal{H}_p$ to quantum Fisher information 
in complete Hilbert space $\mathcal{H} = \mathcal{H}_c 
\otimes \mathcal{H}_p$ for unbounded discrete-time quantum 
walk for 200 time steps and different values of $\theta$. 
The initial state of the walker is $\ket{+} \otimes \ket{x=0}$. }
\label{QFIPositionDComplete}
\end{figure}

%
%
Figure \,\ref{QFIPositionDComplete}  shows the ratio  
$H_w(\theta)/H_f(\theta)$ between the QFI of the walker's position space
and the full QFI. As it is apparent from the plot, after an initial
transient, the ratio saturates to a constant value. More explicitly, this means that performing measurements involving the 
sole position degree of freedom of DTQW provides a considerable information 
about the coin parameter (quantified by $H_w(\theta)$), when compared to the 
full information that it is in principle available (quantified by $H_f(\theta)$).
Notice that by measurements performed on the position degree of freedom we {\em do
not mean} just position measurement (whose performances are investigated in the next Subsection) but rather {\em any} possible measurement on the walker's position Hilbert 
space. 
\par
\par
\begin{figure}[h!]
\includegraphics[width=0.49\columnwidth]{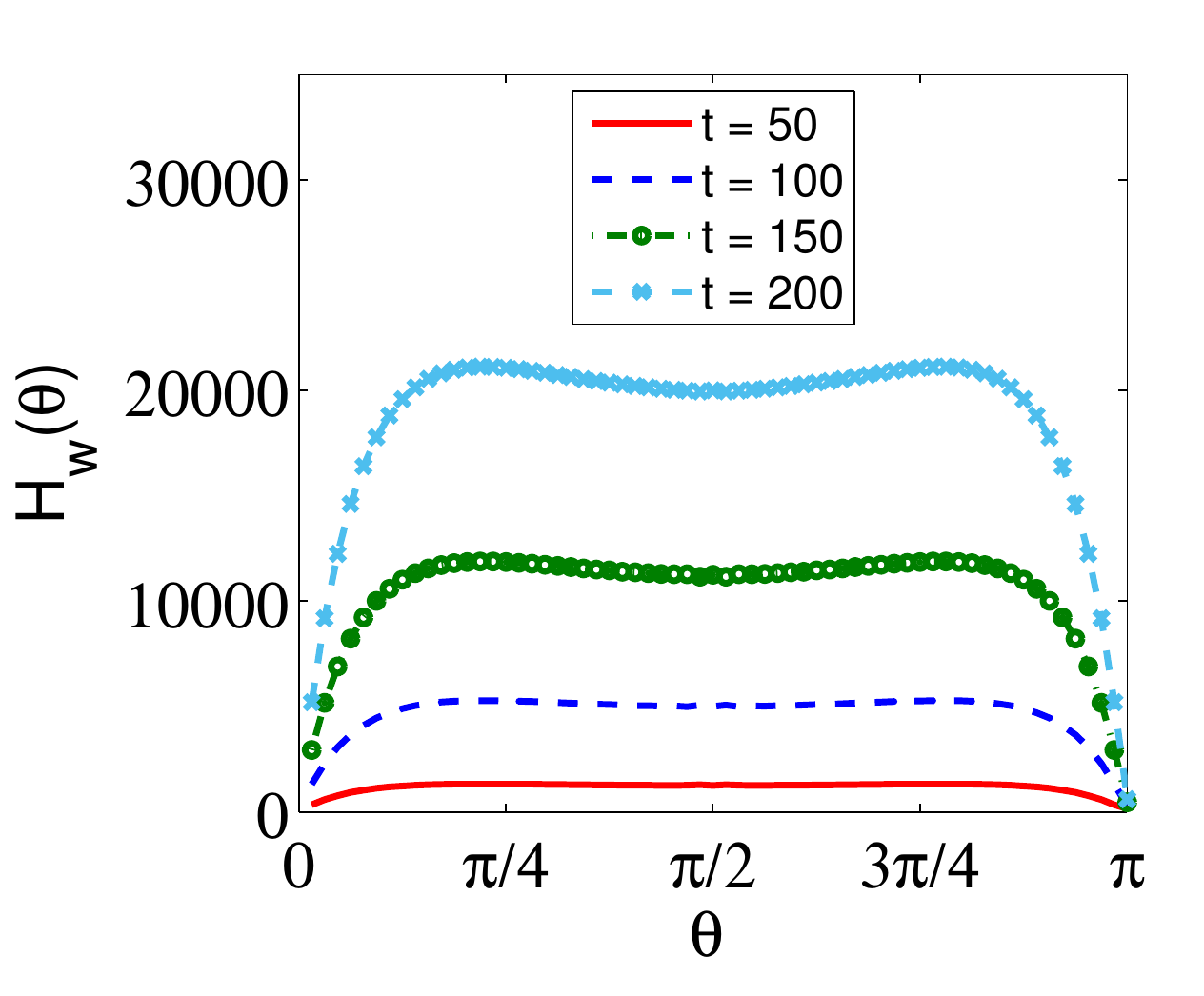} 
\label{fig:QFIUBPVstheta}
\includegraphics[width=0.49\columnwidth]{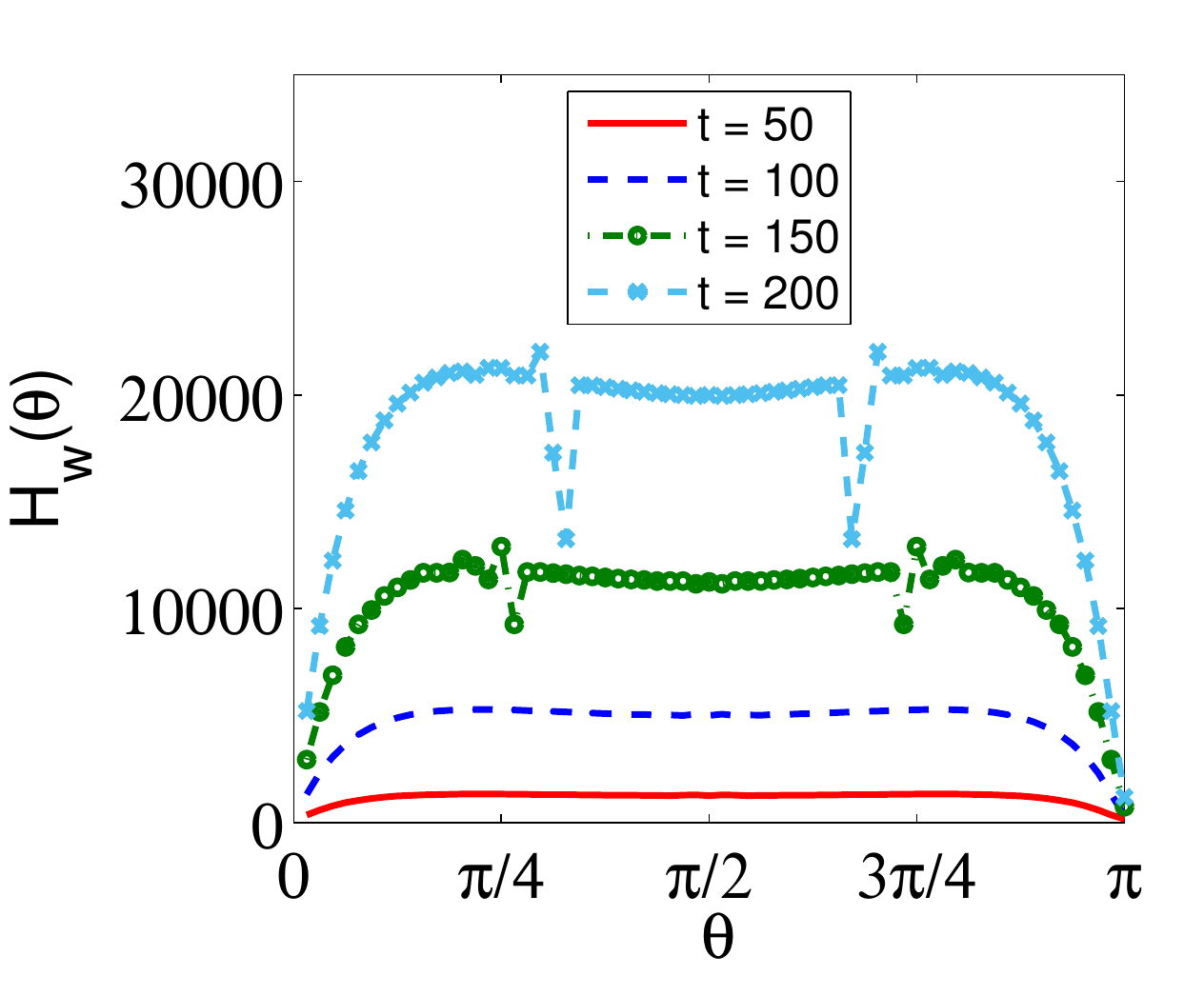} 
\label{fig:QFIBPVstheta}
\caption{The walker's position space QFI $H_w(\theta)$ 
as a function of $\theta$ evaluated 
after a different number of time steps. The unit of $H_w(\theta)$ is inverse square of unit of $\theta$ therefore here it is $(rad)^{-2}$ and unit of $\theta$ is $radian$. The differences 
between the QFI $H_w(\theta)$ of unbounded and bounded 
DTQWs is again due to interference effects in bounded DTQW. 
The initial state of the walker is $\ket{+} \otimes \ket{x=0}$.}
\label{fig:QFIPVstheta}
\end{figure}
\begin{figure}
\includegraphics[width=0.9\columnwidth]{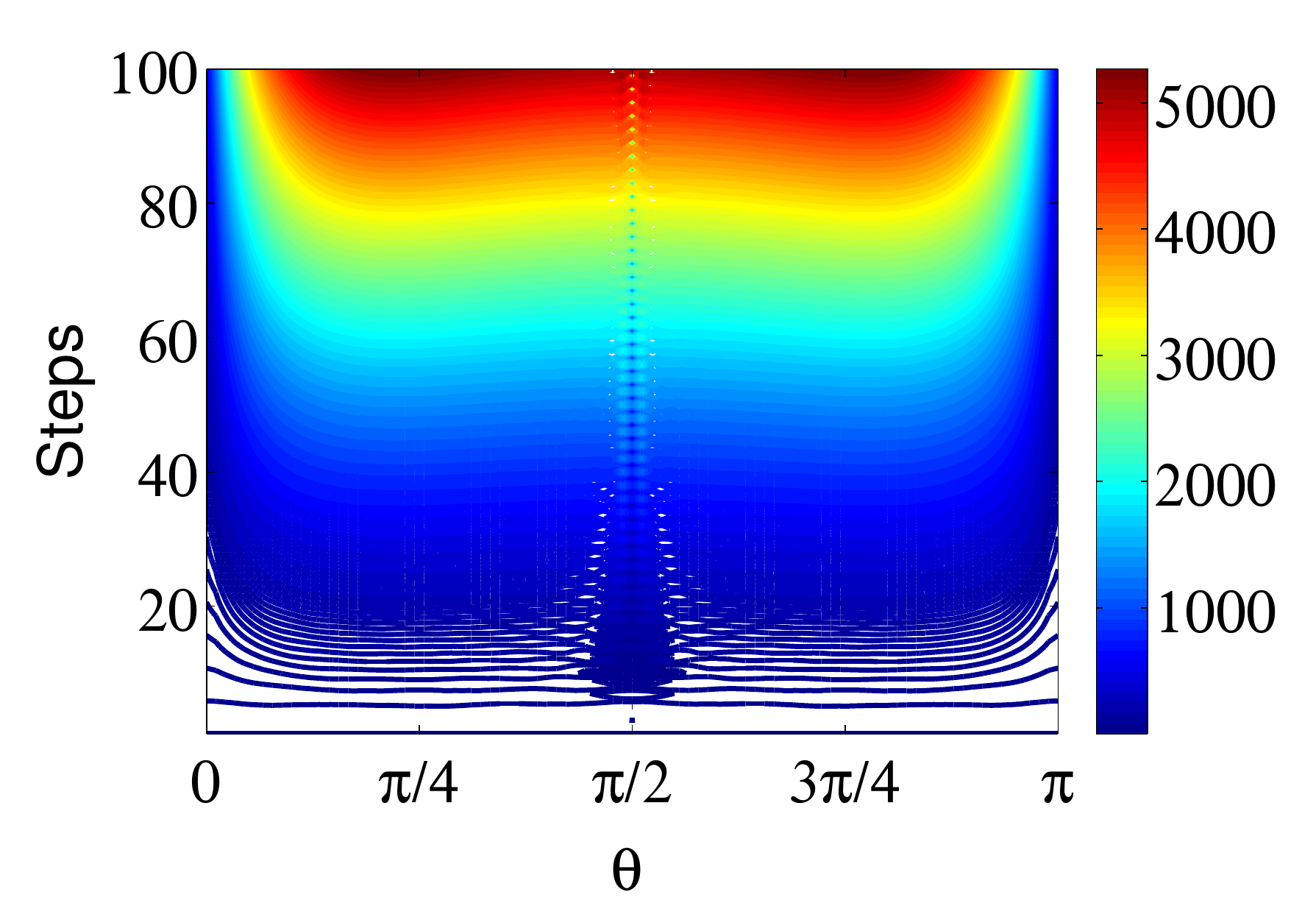}
\caption{QFI in position space ($H_w$) as a function of time steps and $\theta$ for DTQW. The initial state of the walker is $\ket{+} \otimes \ket{x=0}$.}
\label{fig:ContourQFI}
\end{figure}

Figure\,\ref{fig:QFIPVstheta} 
shows the walker QFI $H_w(\theta)$ as a function of $\theta$, for different, fixed, numbers of time steps for both unbounded and bounded DTQWs. It shows that $H_w(\theta)$ increases with $\theta$ initially and than slowly decreases up to $\theta = \pi/2$. 
For $\theta$ ranging from $\theta = \pi/2$ to $\theta = \pi$ the behavior 
is mirrored, because of the symmetry of the quantum coin operation between.  
As it may be seen from the plots the behaviors of the QFIs for unbounded and 
bounded DTQWs are very similar, except for a few more oscillations seen in the bounded
case. In other words, the boundless DTQW is not particularly detrimental for
its use as a probe for the coin parameter. Fig. \ref{fig:ContourQFI} shows QFI in position space $H_w$ of DTQW in one dimension for the estimation of the coin parameter as a function of time step, and coin parameter $\theta$ is shown. This shows that for every coin parameter the QFI in position space increases with time step and therefore probability distribution measurement in position space after a larger time step will give a better estimation of coin parameter $\theta$.

\subsection{\label{subsec3} The FI of walker's position 
measurement in discrete-time quantum walk}
We now turn our attention to the performances of a specific measurement, 
perhaps the most natural one, i.e., the measurement on 
the position of the walker. The conditional probability of finding 
the walker at position $x$ at time $t$, given that the value of the coin parameter
is $\theta$, is given by $p(x|\theta) = \hbox{Tr}[\Pi_x \rho_w (\theta)]$ 
where $\{\Pi_x\}=\{|x\rangle\langle x|\}$ is the set of position projection 
operators, and $\rho_w(\theta)$ is the density matrix of the walker, i.e., the 
statistical operator of Eq. (\ref{rhow}). In other words, the position distribution
of the walker is given by the diagonal elements of the density matrix $\rho_w (\theta)$ 
in the position representation.
\par
Since $\rho_w (\theta)$ is carrying information on $\theta$ at any time, measuring
the position provides information about the value of $\theta$. In order to 
quantify this information, i.e., to quantify how much information about $\theta$
may be obtained by looking at the walker's probability distribution, one has to evaluate the 
position Fisher information using  Eq.\,\eqref{FI}, i.e. 
\begin{align}
F_x(\theta) = \sum_{x} \frac{\left[\partial_{\theta} p(x|\theta)\right]^2}{p(x|\theta)}\,.
\end{align}
According to the quantum Cramer-Rao bound we have $F_x (\theta) \leq H_w (\theta)$,
and, thus, besides the absolute value of $F_x(\theta)$, we are interested in investigating 
how far $F_x (\theta)$ is from its bound  $H_w (\theta)$; i.e., we want to compare 
the information extracted from position measurement to the maximum information 
available measuring the sole walker.
\par
\begin{figure}
\includegraphics[width=0.49\columnwidth]{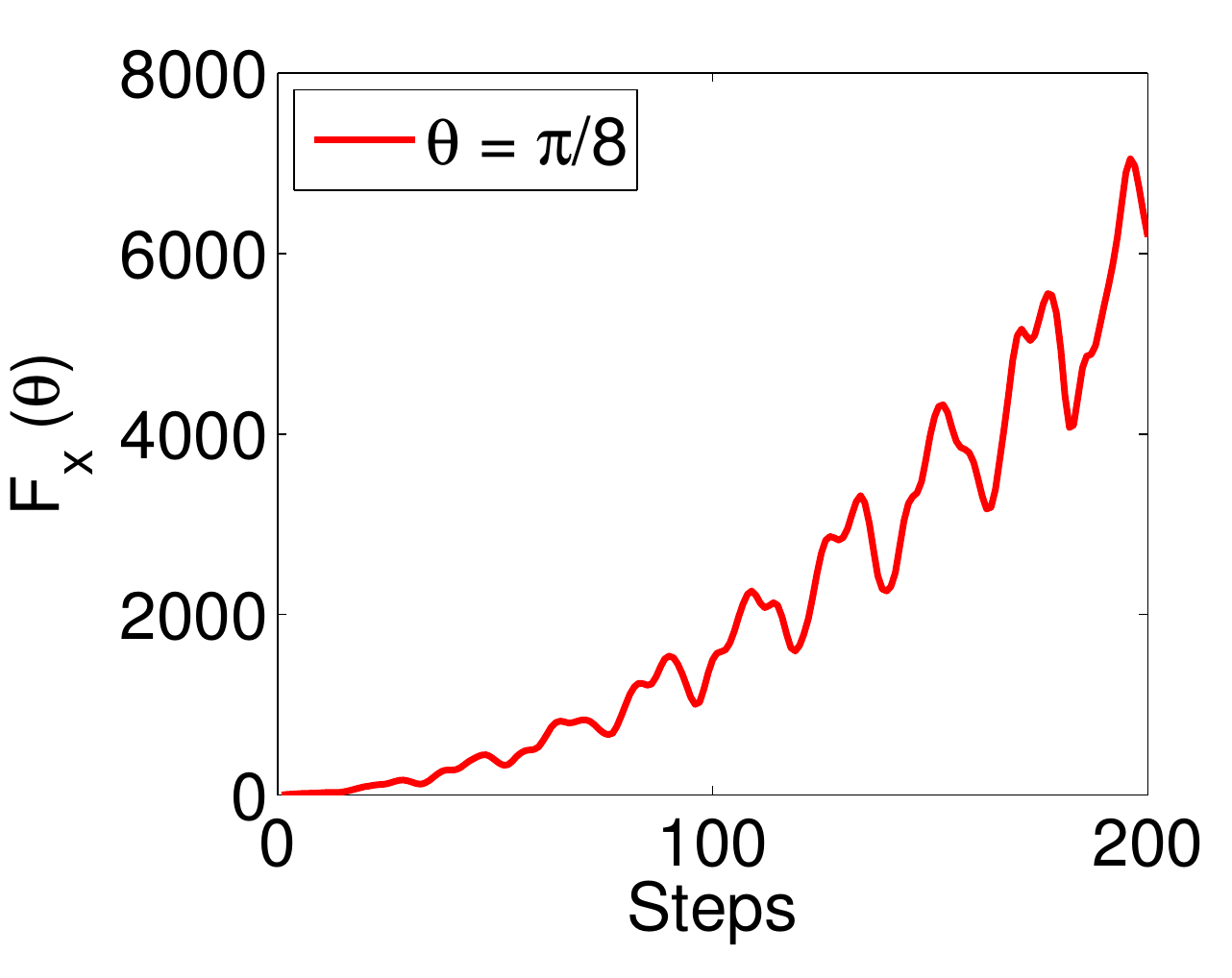}
\includegraphics[width=0.49\columnwidth]{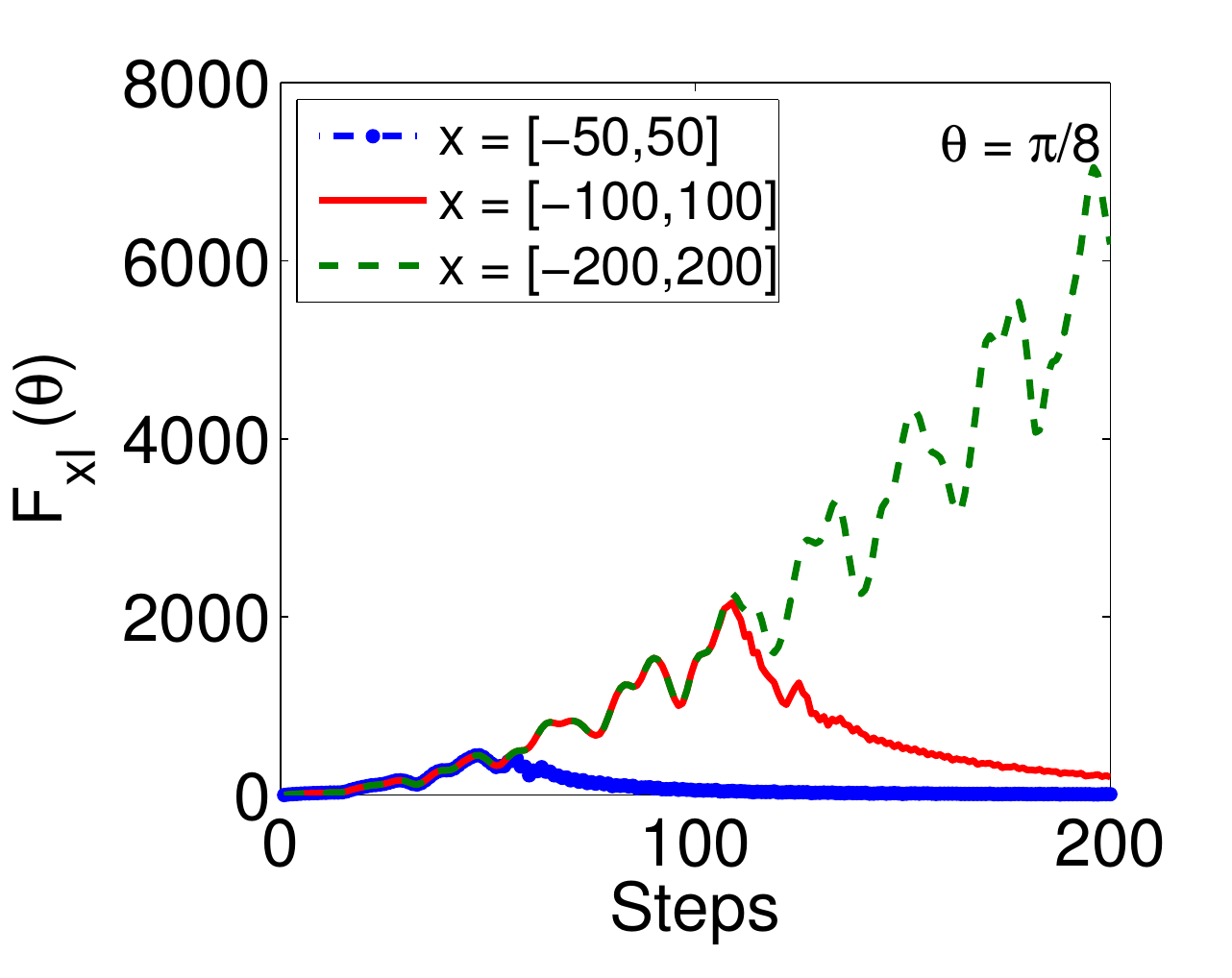}
\includegraphics[width=0.49\columnwidth]{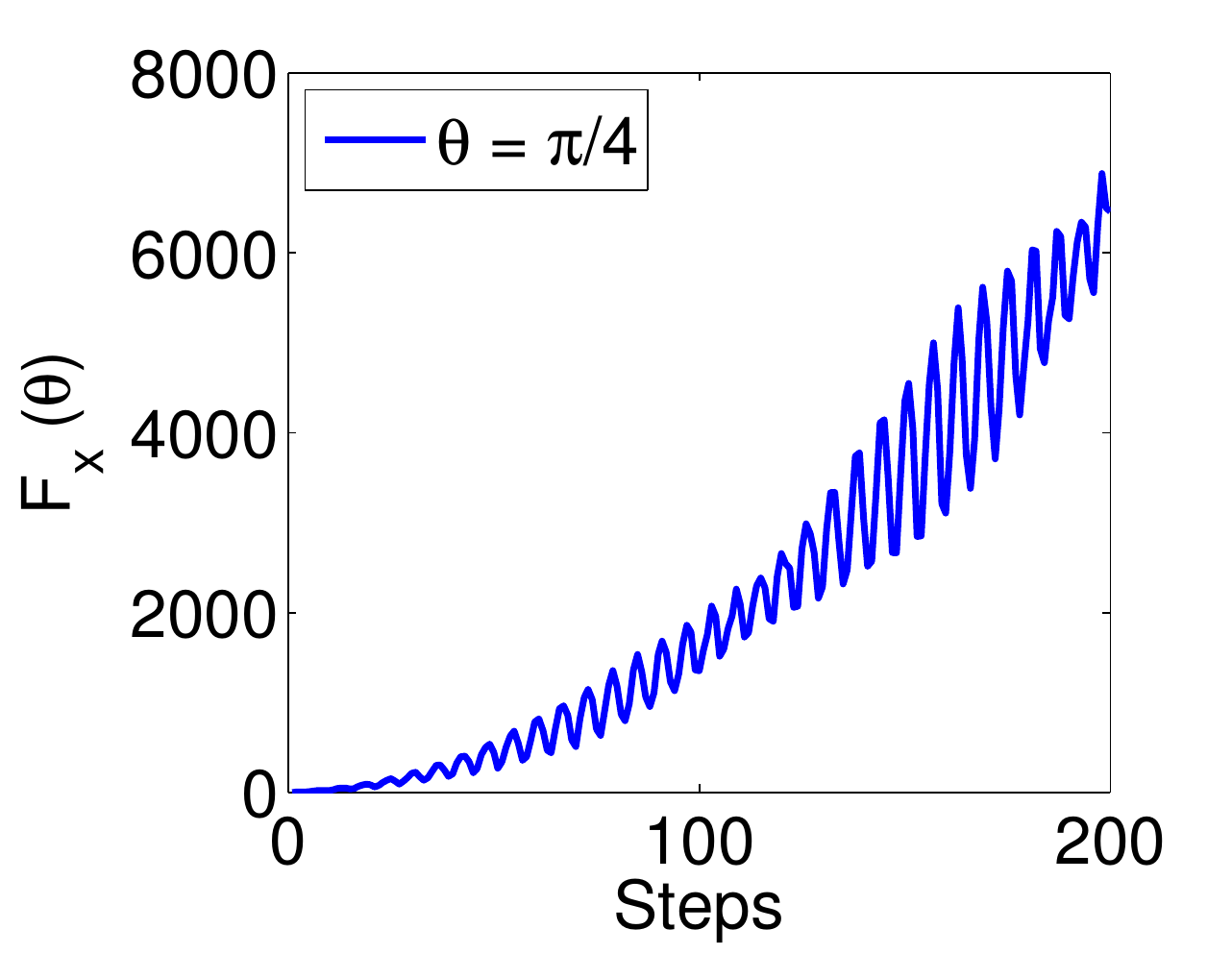}
\includegraphics[width=0.49\columnwidth]{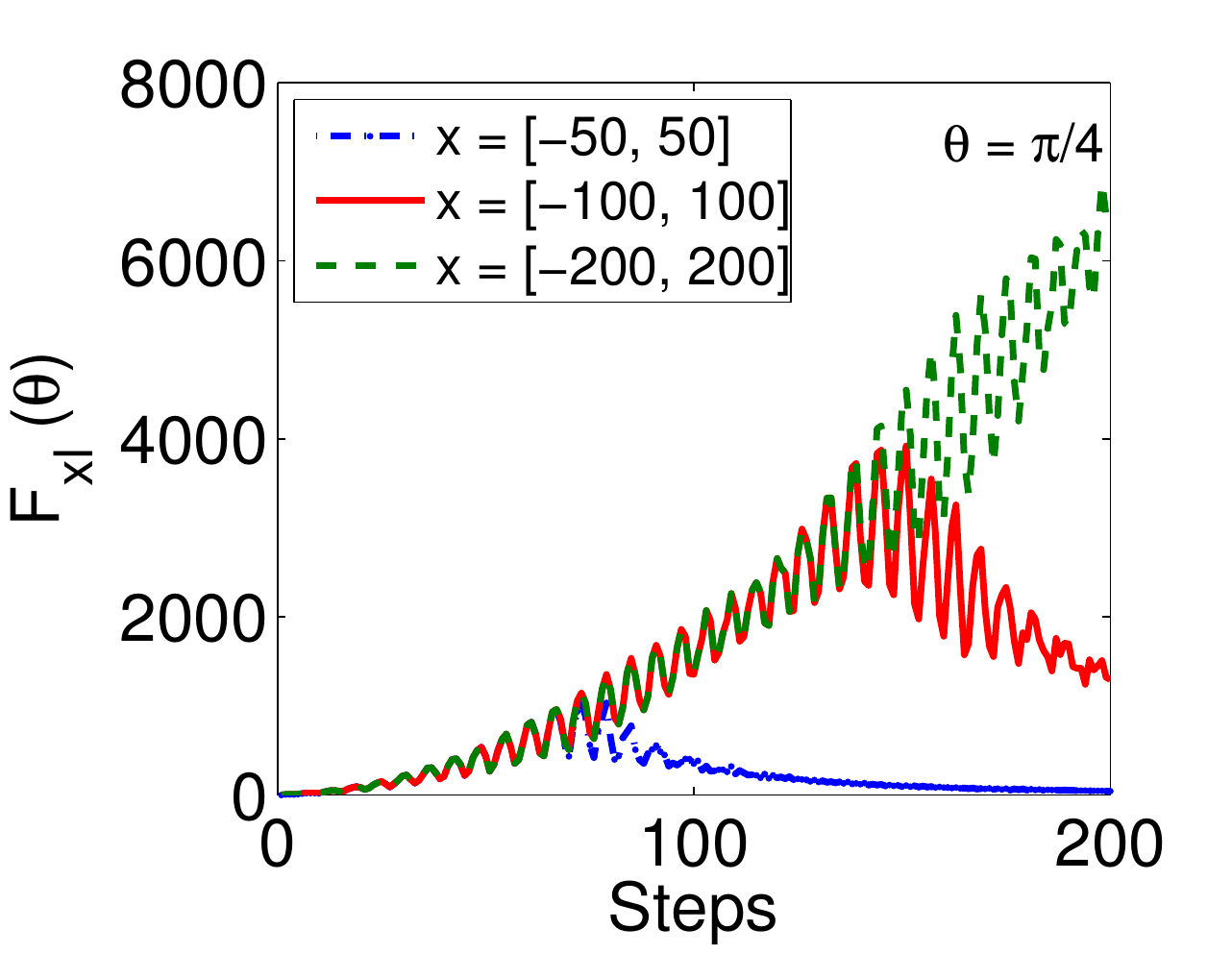}
\includegraphics[width=0.49\columnwidth]{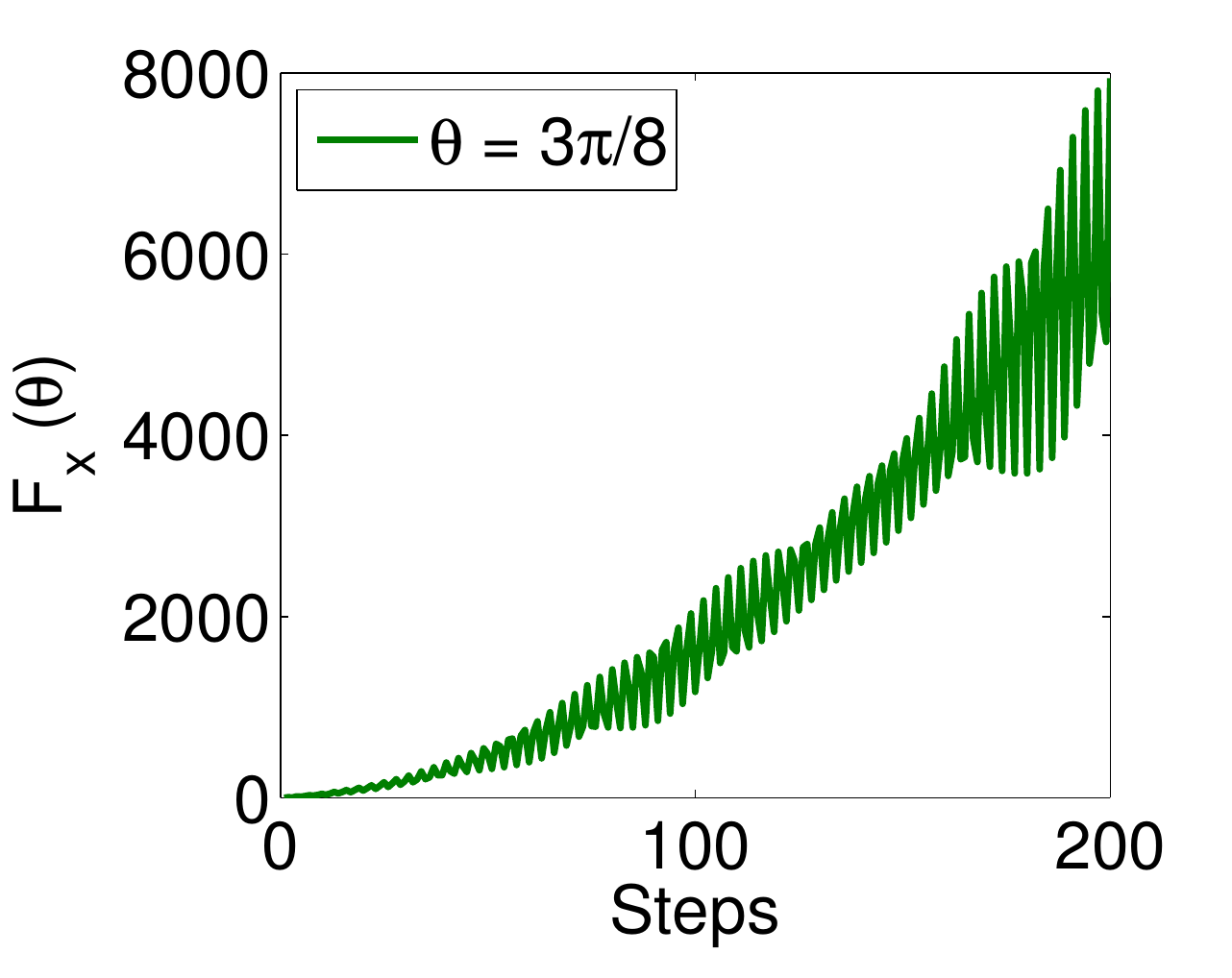}
\includegraphics[width=0.49\columnwidth]{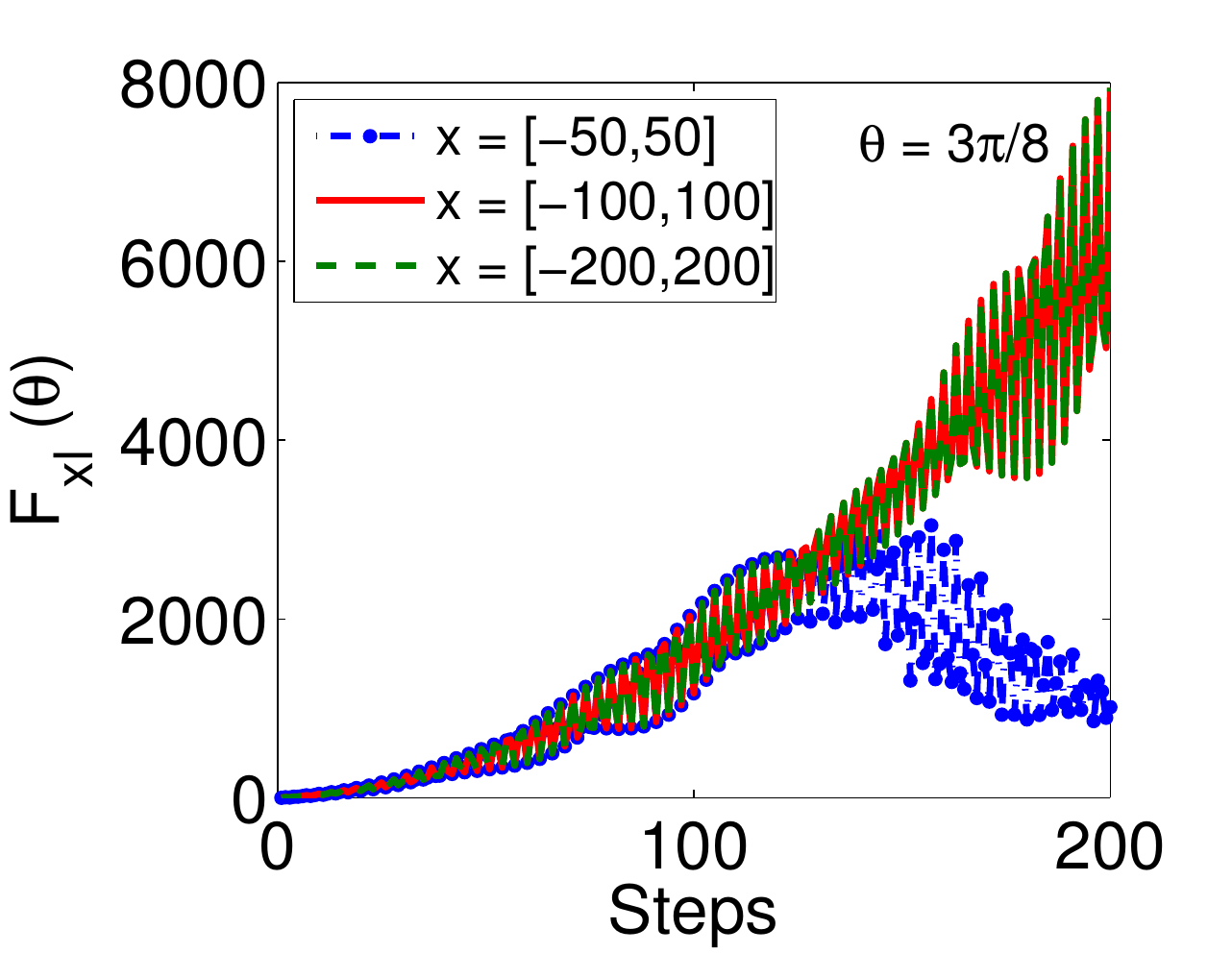}
\caption{The Fisher informations $F_x(\theta)$ (left panels) and 
$F_{xl}(\theta)$ (right panels) as a function of time for unbounded 
DTQW and different values of $\theta$. The unit of $F_{x}(\theta)$ and $F_{xl}(\theta)$ is inverse square of unit of $\theta$ therefore here it is $(rad)^{-2}$ and unit of $\theta$ is $radian$. The initial state of the 
walker is $\ket{+} \otimes \ket{x=0}$. Both sets of plots are for unbounded
DTQW with $F_x(\theta)$ referring to the information
extractable by a full position measurement, whereas 
$F_{xl}(\theta)$ quantifies the information that may be gained by
measurements with limited access to the position of the walker (see text for details). 
The insets of the right panels are legends for the region $S$, accessible 
by the measurement.}
\label{CFI}
\end{figure}

The behavior of $F_x(\theta)$ as a function of time is illustrated in the left
panels of Fig.\,\ref{CFI} for different values of $\theta$. The FI $F_x(\theta)$
oscillates in time, with the envelope increasing as $t^2$, i.e., $F_x(\theta)$ shows
the same scaling as $H_w(\theta)$ and $H_f(\theta)$. The right panels illustrate
instead the behavior of $F_{xl}(\theta)$, which is the Fisher information 
of {\em limited} position measurement, i.e., measurement performed with detectors
not able to access (i.e., to {\em look at}) all the possible walker's sites, but
rather only to a subset $S$, even though the DTQW is defined on an unbounded 
position space. According to Eq.\,\eqref{FI} we have
\begin{align}
F_{xl}(\theta) = \sum_{x\in S} 
\frac{\left[\partial_{\theta} p(x|\theta)\right]^2}{p(x|\theta)}\,,
\end{align}
where the position distribution is still given by $p(x|\theta) = \hbox{Tr}[\Pi_x \rho_w (\theta)]$, however with $x\in S$. In the right panels of Fig. \ref{CFI} we 
show the behavior of $F_{xl}(\theta)$ as a function of time for different 
values of $\theta$ and $S$. The overall message is that for short 
time, when the walker has negligible amplitude to be outside $S$, there 
are little differences between $F_{x}(\theta)$ and $F_{xl}(\theta)$, whereas
for a number of time steps of the order of $|S|$ the walker is {\em walking beyond S} 
and striking differences start to appear. In particular, since in this case the 
measurement is not recording the full position information, the FI $F_{xl}(\theta)$ 
starts to decreases with time.
\par
\par
\begin{figure}[h!]
\includegraphics[width=\columnwidth]{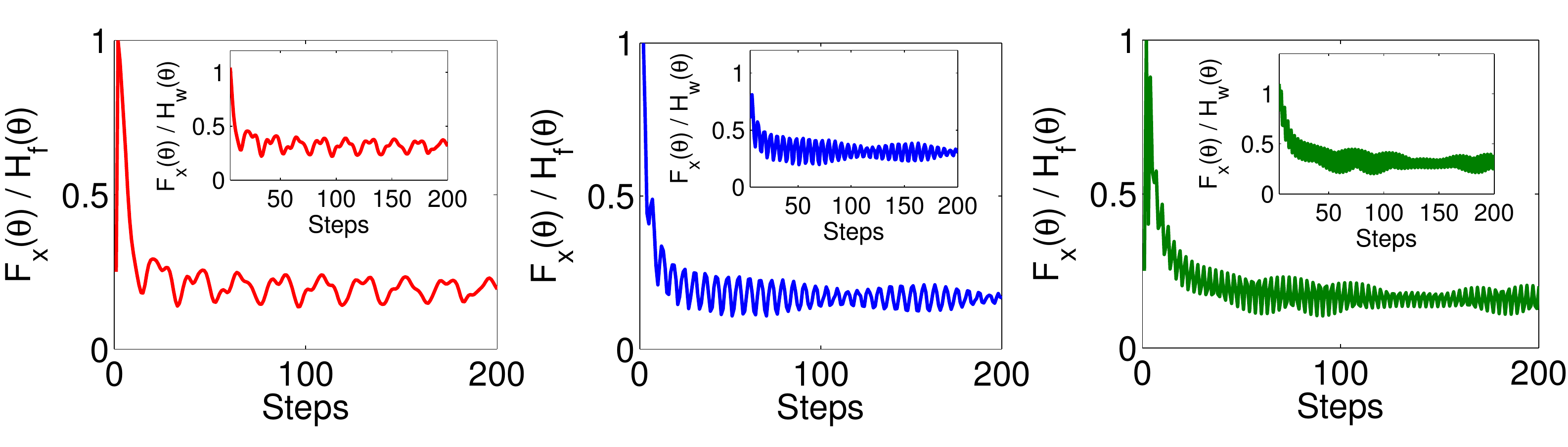}
\caption{The ratio $F_x(\theta)/H_f(\theta)$ between the position Fisher 
information and the full quantum Fisher information as a function of time 
and for different values of $\theta$. The insets show the ratio 
$F_x(\theta)/H_w(\theta)$ between the position Fisher information and 
the walker's position space quantum Fisher information. All the plots refer to unbounded
DTQW. The initial state of the walker is $\ket{+} \otimes \ket{x=0}$.}
\label{CFIComplete}
\end{figure}

In order to assess the overall performances of position measurements 
we consider the two ratios $F_x(\theta)/H_f(\theta)$ and 
$F_x(\theta)/H_w(\theta)$ between the Fisher information of position 
measurement and the full QFI or the walker QFI, respectively. In Fig.\,
\ref{CFIComplete} we show both the ratios as a function of time and for
different values of the coin parameter $\theta$.  
\par

\subsection{\label{subsec4} QFI in split-step quantum walk}
Split-step quantum walk is a special form of discrete-time quantum walk where a single step is split into two half step using two coin operators $C_{\theta_1}$ and $C_{\theta_2}$ and two shift operators $S_-$ and $S_+$. 
Split-step quantum walk has been used to simulate topological insulators \cite{tarasinski2014scattering, asboth2012symmetries, kitagawa2012observation}, Dirac cellular automata \cite{mallick2016dirac}, and Majorana modes and edge states \cite{zhang2017decomposition} where the two coin operations play an important role. It has also been mapped to two period standard discrete-time quantum walks \cite{zhang2017decomposition,kumar2018bounds}.
The evolution operator for split-step quantum walk is given by $U = S_+C_{\theta_2}S_-C_{\theta_1}$ where,
\begin{align}
S_+ &= \sum_{x} (\ket{\uparrow}\bra{\uparrow} \otimes \ket{x}\bra{x} + \ket{\downarrow}\bra{\downarrow} \otimes \ket{x+1}\bra{x})  \\
S_- &= \sum_{x} (\ket{\uparrow}\bra{\uparrow} \otimes \ket{x-1}\bra{x} + \ket{\downarrow}\bra{\downarrow} \otimes \ket{x}\bra{x})
\end{align}
and the coin operator is given by,
\begin{align}
C_{\theta_j} = \begin{pmatrix}
\cos\theta_j & -i \sin\theta_j \\
-i \sin\theta_j & \cos\theta_j
\end{pmatrix} \otimes \sum_{x} \ket{x}\bra{x}, \nonumber \\
\end{align}
where $j = 1,2$.
\par
\begin{figure}[h!]
\includegraphics[width=0.9\columnwidth]{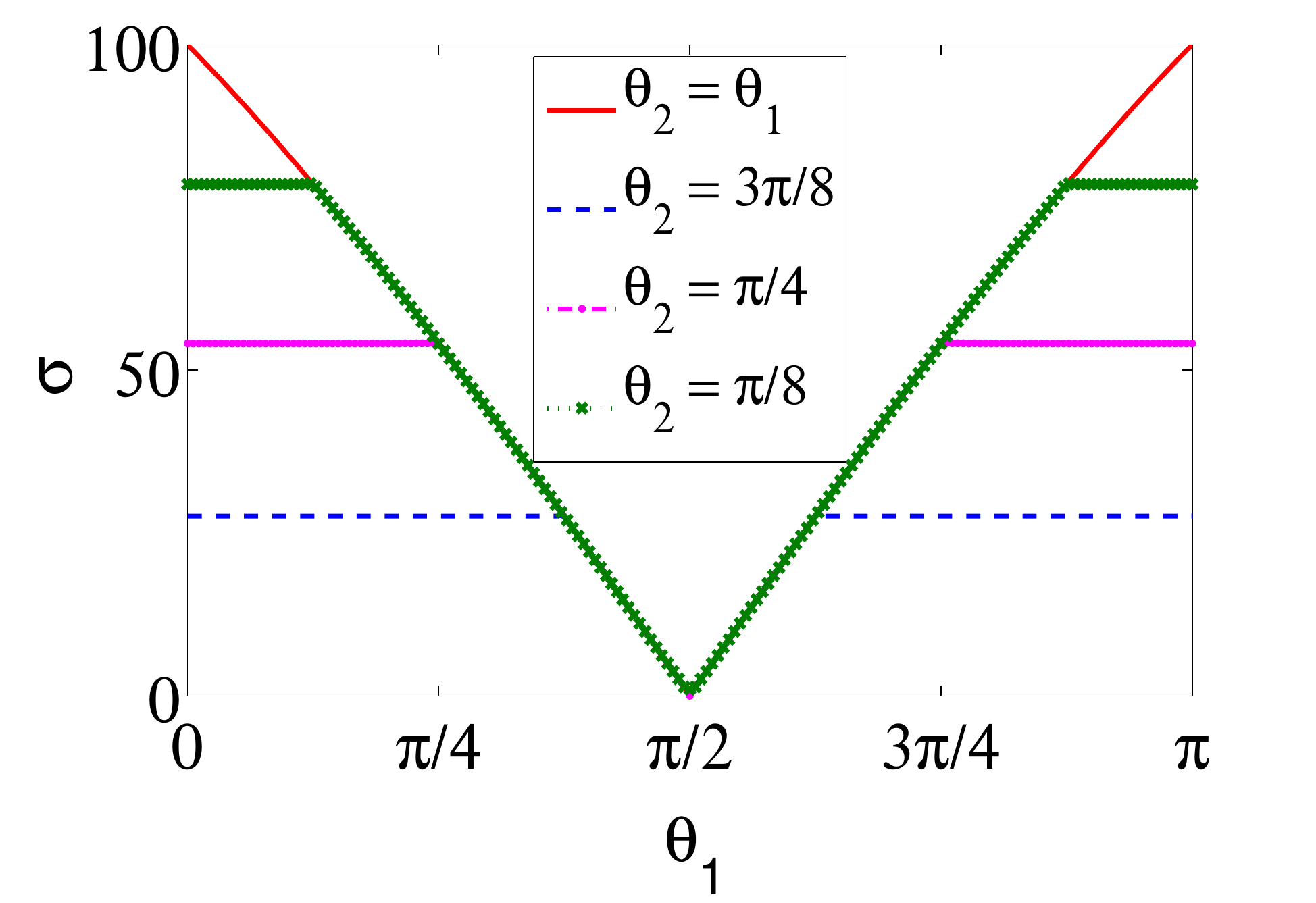}
\caption{The standard deviation for split-step quantum walk as function of  $\theta_1$ for different values of $\theta_2$ after 100 steps of walk. The unit of $\theta_1$ and $\theta_2$ is $radian$. Standard deviation is always bounded by the larger $\theta$ parameter. Therefore it only gives information of the evolution parameter with the higher value. The initial state of the walker is $\ket{+} \otimes \ket{x=0}$.}
\label{STD_SSQW}
\end{figure}
\par
\begin{figure}[h!]
\includegraphics[width=\columnwidth]{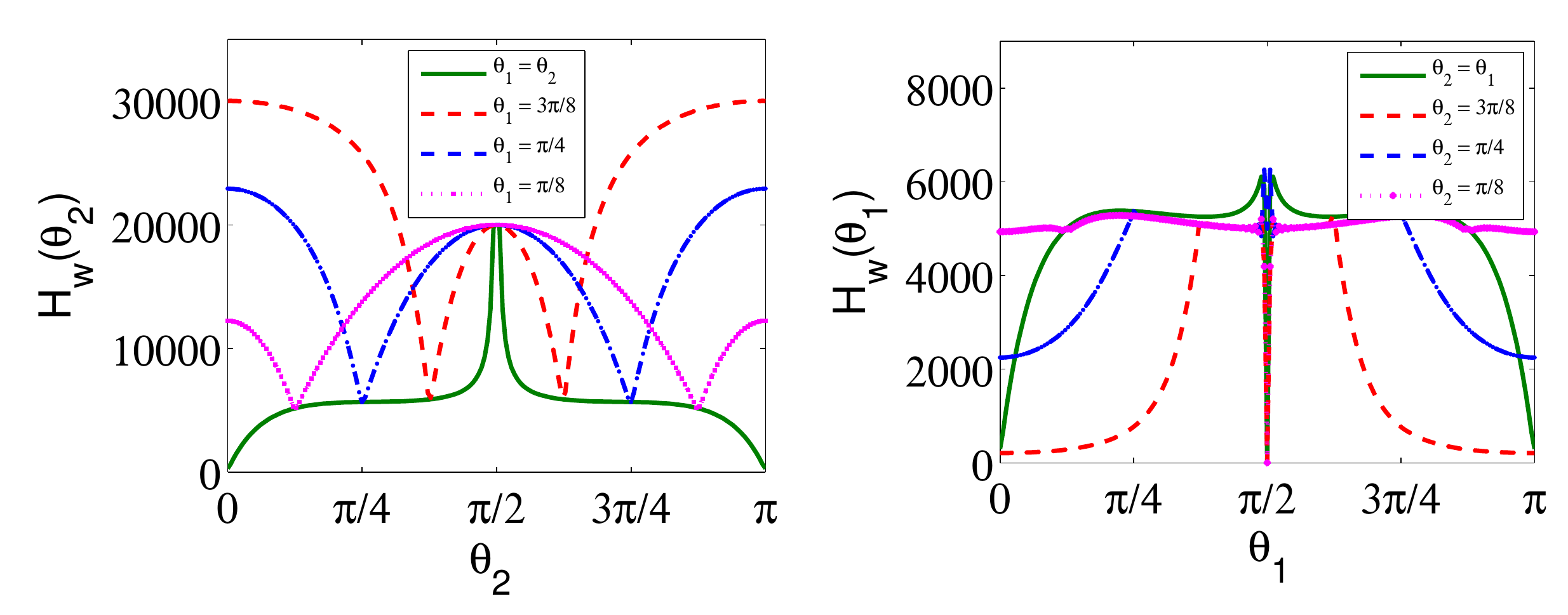}
\caption{Quantum Fisher information in position space with respect to the evolution parameters $\theta_2$ (left) and $\theta_1$ (right) when other parameters are fixed. The unit of $H_w(\theta)$ is inverse square of unit of $\theta$ therefore here it is $(rad)^{-2}$ and unit of $\theta$ is $radian$. Having clearly distinct plot for both parameters after 100 time steps helps us to uniquely probe both parameters independently.  The initial state of the walker is $\ket{+} \otimes \ket{x=0}$.}
\label{QFI_SSQW}
\end{figure}

Estimation of both the parameters $\theta_1$ and $\theta_2$ using standard deviation is not possible, as has already been studied in the past \cite{kumar2018bounds}. In Fig. \ref{STD_SSQW} we show the standard deviation of unbounded split-step quantum walk after 100 steps of walk as a function of $\theta_1$ when $\theta_2$ is fixed. We can note that the standard deviation is always bounded by the larger of the two parameters $\theta_1$ and $\theta_2$. But using the quantum Fisher information for both the parameters individually in position space, each of the parameters can be estimated. It can be seen in Fig. \ref{QFI_SSQW} that QFI in position space with respect to $\theta_2$ for different values of $\theta_1$ shows that $H_w(\theta_2)$ is minimum for $\theta_2 = \theta_1$ and QFI in position space with respect to $\theta_1$ for different values $\theta_2$ shows that $H_w(\theta_1)$ is maximum for $\theta_2 = \theta_1$. Since QFI in position space is a measure of how precisely one can estimate the evolution parameters on measurement of probability distribution in position space, Fig. \ref{QFI_SSQW} shows that, given the value of parameter $\theta_1$, $\theta_2$ can be estimated more precisely when $\theta_2 \neq \theta_1$ as $H_w(\theta_2)$ is minimum when $\theta_2 = \theta_1$. Similarly $H_w(\theta_1)$ shows that the amount of information of $\theta_1$ is maximum when $\theta_1 = \theta_2$ on measurement of the probability distribution. This estimation is not possible by just measuring the standard deviation in the split-step quantum walk. 
\par

\section{\label{sec4}Conclusion}
In this paper, we have investigated probing techniques for the coin 
parameter $\theta$ of discrete-time quantum walk, which, in turn, plays 
a crucial role in providing quadratic speed-up over its classical 
counterpart. In particular, we have addressed the ultimate bounds to 
precision, as obtained by performing the optimal measurement on the 
particle. 
Our approach is based on the fact that the walker's coin space 
entangles with the position space after the very first step 
of the evolution, such that we may estimate the value of the 
coin parameter $\theta$ by performing measurements on the sole 
position space of the walker. 
\par
We have found that the QFI of the walker's position space $H_w(\theta)$ increases 
with $\theta$ and with time which, in turn, may be seen as a metrological 
resource. We also find a difference in the QFI of {\em bounded} 
and {\em unbounded} DTQWs, and provide an interpretation of the 
different behaviors in terms of interference in the position 
space. We have also compared $H_w(\theta)$ to the full 
QFI $H_f(\theta)$, i.e., the QFI of the walker's position 
plus coin state, and find that their ratio is dependent on 
$\theta$, but saturates to a constant value, 
meaning that the walker may probe its coin parameter quite 
faithfully.
Finally, we have found that if one has access to a limited 
region in position space, the QFI depends only on the sites 
with non-zero probability of finding a particle. Therefore, when 
one has access to an incomplete position space, after some steps 
(equal to half of the number of accessible sites) we see 
a decrease of QFI.
\par
Though standard deviation and group velocity help us to estimate one-parameter QW, they fail to provide a reasonable estimation in the case of bounded QW and two-parameter split-step QW. We can overcome this using QFI. 
Our results show that estimation of the coin parameter in DTQW is 
possible with realistic detection schemes and pave the way 
for further developments in the field of quantum probing for
complex networks.
\section*{Acknowledgment}
C.M.C. would like to thank the Department of Science and Technology, 
Government of India for the Ramanujan Fellowship Grant 
No. SB/S2/RJN-192/2014. This work has been supported by SERB 
through Project No. VJR/2017/000011. M.G.A.P. is a member of GNFM-INdAM.  
\bibliography{qfi_dtqw}
\bibliographystyle{ieeetr}
\end{document}